\documentclass[twocolumn]{aastex62}

\usepackage{enumitem}
\usepackage{mathtools}

\shorttitle{Partial, zombie, and full TDEs}
\shortauthors{Nixon et al.}

\begin{document}

\title{Partial, zombie, and full tidal disruption of stars by supermassive black holes}

\correspondingauthor{C.~J.~Nixon}
\email{cjn@leicester.ac.uk}

\author[0000-0002-2137-4146]{C.~J.~Nixon}
\affiliation{Department of Physics and Astronomy, University of Leicester, Leicester, LE1 7RH, UK}

\author[0000-0003-3765-6401]{E.~R.~Coughlin}
\affiliation{Department of Physics, Syracuse University, Syracuse, NY 13244}

\author[0000-0003-1354-1984]{P.~R.~Miles}
\affiliation{Department of Physics, Syracuse University, Syracuse, NY 13244}

\begin{abstract}
We present long-duration numerical simulations of the tidal disruption of stars modelled with accurate stellar structures and spanning a range of pericentre distances, corresponding to cases where the stars are partially and completely disrupted. We substantiate the prediction that the late-time power-law index of the fallback rate $n_{\infty} \simeq -5/3$ for full disruptions, while for partial disruptions---in which the central part of the star survives the tidal encounter intact---we show that $n_{\infty} \simeq -9/4$. For the subset of simulations where the pericenter distance is close to that which delineates full from partial disruption, we find that a stellar core can reform after the star has been completely destroyed; for these events the energy of the zombie core is slightly positive, which results in late-time evolution from $n \simeq -9/4$ to $n \simeq -5/3$. We find that self-gravity can generate an $n(t)$ that deviates from $n_{\infty}$ by a small but significant amount for several years post-disruption.  In one specific case with the stellar pericenter near the critical value, we find self-gravity also drives the re-collapse of the central regions of the debris stream into a collection of several cores while the rest of the stream remains relatively smooth. We also show that it is possible for the surviving stellar core in a partial disruption to acquire a circumstellar disc that is shed from the rapidly rotating core. Finally, we provide a novel analytical fitting function for the fallback rates that may also be useful in a range of contexts beyond TDEs.
\end{abstract}


\section{Introduction}
The tidal disruption of one object by another is a recurring theme in astrophysics. For example, comets can be tidally disrupted by planets (e.g., Comet Shoemaker-Levy 9; \citealt{Chapman:1993aa}) and asteroids can be tidally disrupted by white dwarfs \citep{Jura:2003aa}; star clusters and small galaxies can be disrupted by larger ones \citep[e.g.][]{Penarrubia:2009aa}; and neutron star mergers, in which each star tidally disrupts the other, are some of the most luminous events in the Universe \citep[e.g.][]{Faber:2012aa}. In this paper we focus on the tidal disruption of stars by supermassive black holes, referred to as tidal disruption events (TDEs; e.g., \citealt{Rees:1988aa}). These events produce luminous, multiwavelength flares that form and evolve on timescales of days to years, which allows one to explore (for example) dormant black holes in galaxy centres, nuclear galactic dynamics, stellar structure, accretion and jet physics. Observations of TDEs across different wavelengths and at various epochs have now been made (e.g., \citealt{Bade:1996aa, Komossa:1999aa, Esquej:2007aa, Gezari:2009aa, Gezari:2012aa, Holoien:2014aa, Miller:2015ab, Vinko:2015aa, Alexander:2016aa, Cenko:2016aa, Holoien:2016aa, Kara:2016aa, vanVelzen:2016aa, Alexander:2017aa, Blanchard:2017aa, Brown:2017aa, Gezari:2017aa, Hung:2017aa, Saxton:2017aa, Brown:2018aa, Pasham:2018aa, Blagorodnova:2019aa, Hung:2019aa, Holoien:2019aa, Leloudas:2019aa, Nicholl:2019aa, Pasham:2019aa, Saxton:2019aa, Hung:2020aa, Hung:2020ab, Holoien:2020aa, Jonker:2020aa, Kajava:2020aa, Li:2020aa, Hinkle:2021aa, Payne:2021aa, vanVelzen:2021aa}); see also the recent reviews by \citet{Alexander:2020aa, vanVelzen:2020aa, Gezari:2021aa}.

The tidal radius $R_{\rm t}$---the distance from the black hole at which the star is expected to be destroyed by tides---is typically estimated by equating the mean self-gravitational force of the star to the gravitational tidal force from the supermassive black hole; the result is 
\begin{equation}
\label{rt}
R_{\rm t} \approx \left(\frac{M_{\bullet}}{M_\star}\right)^{1/3}R_\star
\end{equation}
where $M_{\bullet}$ is the mass of the black hole and $M_\star$ and $R_\star$ are the stellar mass and radius respectively. The impact parameter of the stellar orbit around the black hole is defined as $\beta \equiv R_{\rm t}/R_{\rm p}$, where $R_{\rm p}$ is the pericentre radius of the stellar orbit. Grazing encounters with $\beta \lesssim 1$ typically result in partial disruptions, with some fraction of the star surviving the tidal encounter intact, while deep encounters with $\beta \gg 1$ yield complete disruption of the star \citep[see, for example, the simulation results in][]{Guillochon:2013aa}. The precise value of $\beta$ at which the star is completely disrupted depends on the stellar properties. For example, \cite{Guillochon:2013aa} (see also \citealt{Mainetti:2017aa}) showed that when {a solar-like (i.e., one with a solar mass and radius)} star is modelled as a polytrope, the critical impact parameter for full disruption is $\beta_{\rm c} \simeq 2\,(0.9)$ for $\gamma=4/3\,(5/3)$. In some cases polytropes provide an excellent description of the density profile of a star (see, e.g., Fig.~5 of \citealt{Golightly:2019ab}), but in many cases a polytrope cannot provide a good fit throughout the star. Using accurate stellar structures derived from the {\sc mesa} stellar evolution code \citep{Paxton:2011aa, Paxton:2013aa, Paxton:2015aa, Paxton:2018aa, Paxton:2019aa}, \cite{Golightly:2019ab} show that only 4 out of 9 simulated stars were fully disrupted when $\beta = 3$.

Due to the complexity of the disruption process and the subsequent evolution of the debris stream, numerical simulations are often used to make progress in understanding TDE dynamics. In some cases these simulations have focussed on the initial disruption and debris energy distribution \citep{Lodato:2009aa,Guillochon:2013aa}, the stream dynamics \citep{Coughlin:2015aa,Coughlin:2016ab,Coughlin:2016aa}, disc formation from eccentric orbits \citep{Hayasaki:2013aa,Hayasaki:2016aa,Shiokawa:2015aa,Bonnerot:2016aa}, the impact of stellar rotation \citep{Golightly:2019aa,Sacchi:2019aa}, the importance of stellar structure \citep{Golightly:2019ab,Law-Smith:2019aa} and the impact of general relativity \citep{Gafton:2015aa, Sadowski:2016aa,Tejeda:2017aa, Gafton:2019aa, Andalman:2020aa, Curd:2021aa}.\footnote{The role of magnetic fields in TDEs have also been explored \citep[e.g.][]{Guillochon:2017aa,Bonnerot:2017aa}. For the initial disruption and stream evolution anomalously large field strengths must be used to generate even a minor impact on the results. The role of magnetic fields in TDE disc formation, evolution and the production of jets has not yet been fully established (see, for example, \citealt{Sadowski:2016aa}).}

One of the goals of TDE simulations is to determine the fallback rate: the rate at which the disrupted stellar debris returns to pericentre. This rate is a determining factor in the production of accretion luminosity from the black hole.\footnote{If the returning debris accretes onto the black hole sufficiently rapidly, then the bolometric luminosity is approximately given by $\eta{\dot M}_{\rm fb}c^2$, where $\eta\approx 0.1$ is the accretion efficiency and ${\dot M}_{\rm fb}$ is the fallback rate. However, any significant delays in the circularisation of the returning debris stream or the subsequent accretion of matter through the disc may invalidate this simple relation. \cite{Mockler:2019aa} find that any delays measured from observed TDEs are small.} For matter to accrete on to the black hole, or be expelled in winds/jets (e.g., \citealt{Strubbe:2009aa, Coughlin:2014aa, Metzger:2016aa}), energy must be extracted from the debris orbits, and some (perhaps most) of this energy is released as radiation (for a discussion of energy release in TDEs see \citealt{Lu:2018aa}). The classical prediction for TDEs is that the power-law index of the fallback rate is $-5/3$. This was derived by \citet[][see \citealt{Phinney:1989aa}]{Rees:1988aa} by assuming that the stellar debris follows Keplerian orbits (with fixed orbital energy $\epsilon$) around the black hole and that for most of the debris the mass-energy distribution (${\rm d}M/{\rm d}\epsilon$) would be ``flat" (i.e., independent of $\epsilon$). Subsequently, \cite{Lodato:2009aa} showed that a more realistic stellar density profile (i.e., one that is not constant) implies that the power-law index of the fallback rate only reaches a value close to $-5/3$ at late (but still observable) times and after a significant amount of the bound debris has been accreted. Similar results were found by, for example, \cite{Guillochon:2013aa} and \cite{Golightly:2019ab}, the latter showing that for a solar-like star modelled as a polytrope disrupted by a $10^6 M_\odot$ black hole the fallback rate power-law index $n(t)$ remains shallower than $-5/3$ with $-1.6 < n(t) < -1.4$ for $0.25 < t/{\rm yr} < 1.25$.

More recently we have shown that the late-time fallback rate power-law index ($n_\infty$) for a {\it partial} TDE---in which the star is not fully disrupted and some of the star remains intact---is $\approx -9/4$, and not $-5/3$ \citep{Coughlin:2019aa}. This value is almost completely independent of the mass fraction of the surviving core, $\mu = M_{\rm c}/M_{\bullet}$, where $M_{\rm c}$ is the mass of the surviving core and $M_{\bullet}$ is the mass of the black hole (see Equation 14 of \citealt{Coughlin:2019aa}). This arises from the fact that the debris that returns at late times originates asymptotically close to the Hill's radius---the region surrounding the core in which the core's gravity dominates over that of the black hole---of the surviving core, and thus is always affected by the core's gravity for any (non-zero) mass. In \cite{Miles:2020aa} we presented numerical simulations of partial TDEs employing polytropes to model the stars, and recovered the $\propto t^{-9/4}$ power-law for the simulations which left a bound core, substantiating the predictions of \cite{Coughlin:2019aa}. 

In this paper we extend the work of \cite{Miles:2020aa} by simulating TDEs with several different stellar models with accurate density profiles, and a larger range of impact parameters ($\beta \equiv R_{\rm t}/R_{\rm p}$) to explore the transition from $-5/3$ to $-9/4$. The numerical simulations are performed with stars on parabolic orbits with respect to the central black hole. We use some of the stellar models employed by \cite{Golightly:2019ab}, which were calculated with the {\sc mesa} stellar evolution code. With these simulations we are able to clearly identify a dichotomy in the late-time power-law index of the fallback rate, with each case producing indices that are either close to $-5/3$ or $-9/4$ depending on whether the star is fully or only partially disrupted. We find some cases where there is a small offset from these values that persists for the duration of the simulations, which we argue is caused by the effects of self-gravity acting along the debris stream. We find that the relationship between the properties of the fallback curves and the stellar structure can be complex, with trends observed for one star no longer present for another star. We examine the fallback rates for full disruptions with $\beta$ up to $\simeq 2\beta_{\rm c}$, where $\beta_{\rm c}$ is the critical $\beta$ for full disruption, finding that self-gravity maintains an important role in establishing the tidally disrupted stream structure. We also find an example of a partial disruption in which the surviving core acquires a circumstellar disc of material, which may provide interesting observable consequences.

The layout of the paper is as follows. In Section~\ref{sims} we describe the numerical simulations and the results. In Section~\ref{fits} we present analytical fits to the fallback data from the simulations. We present our conclusions in Section~\ref{conclusions}. 

\section{Simulations}
\label{sims}
As discussed above there are a wide range of TDE simulations available in the literature. Historically TDEs have been approached with Lagrangian based methods, e.g. Smoothed Particle Hydrodynamics (SPH; \citealt{Gingold:1977aa,Lucy:1977aa}), see for example \cite{Nolthenius:1982aa,Nolthenius:1983aa,Bicknell:1983aa,Evans:1989aa}. This choice is presumably due to the inherent Lagrangian nature of the problem. In recent years breakthroughs in understanding the dynamics of TDEs have been made using both Eulerian (grid-based; e.g. \citealt{Guillochon:2013aa}, \citealt{Sadowski:2016aa} \citealt{Jiang:2016aa}) and Lagrangian (particle-based; \citealt{Lodato:2009aa}, \citealt{Coughlin:2015aa}, \citealt{Golightly:2019ab}) methods.\footnote{There is also still room for progress through traditional ``pen and paper" research \citep[e.g.,][]{Coughlin:2019aa}.} In this work we are primarily interested in the long-term evolution of the debris stream, and thus we choose to use the SPH method, specifically the publicly available SPH code {\sc phantom} \citep{Price:2018aa}, which we have used for previous TDE investigations \citep[see, for example,][]{Coughlin:2015aa,Golightly:2019ab,Golightly:2019aa,Miles:2020aa}.

For the numerical simulations here we set up the stars to have a density profile taken from the outputs of the {\sc mesa} stellar evolution code. Specifically we use the $1M_\odot$ ZAMS, $1M_\odot$ MAMS and $0.3M_\odot$ MAMS models presented in \cite{Golightly:2019ab}.\footnote{ZAMS refers to zero-age main sequence, and here MAMS refers to middle-age main sequence which \cite{Golightly:2019ab} define as being when the stellar core's hydrogen fraction drops below 0.2.} We take the supermassive black hole to have a mass of $10^6 M_\odot$ and model its gravitational field as Newtonian\footnote{For the parameters we simulate here the inclusion of GR effects is not expected to have a strong effect on the disruption process \citep{Gafton:2015aa}. However, GR effects are, of course, necessary for an accurate understanding of the subsequent disc formation and evolution that follows the fallback of the debris stream.}. We take each stellar orbit to be parabolic and initiate its approach to the black hole at a distance of $5R_{\rm t}$ (at which the tidal field from the black hole is a factor of $\approx 125$ weaker than at the tidal radius) and we vary the impact parameter $\beta$. The critical impact parameters are $\beta_{\rm c} \simeq 1.6$, $3.7$, and $1.5$ for the $1M_\odot$ ZAMS, $1M_\odot$ MAMS and $0.3M_\odot$ MAMS stars respectively. For each star we simulate a range of impact parameters with $\beta_{\rm min} < \beta_{\rm c} < \beta_{\rm max}$, and we go from the minimum to maximum values with a small increment of 0.1. As this results in a large set of simulations, we employ moderate resolution with $10^6$ particles for the star. For the range of impact parameters we simulate this is sufficient to model the disruption and stream dynamics. However, we note that for very weak encounters (not simulated here) in which only a small fraction of the star's mass is tidal stripped there is typically only a small number of particles in the resulting debris stream and thus, when analysing such simulations, a larger number of particles in the initial star would be necessary. 

As mentioned above, in a number of our simulations a stellar core survives the tidal encounter (or reforms later; see below). In these cases the hydrodynamical time step in the simulation is vastly reduced compared to core-less simulations owing to the relatively high central pressure and density of the core. Therefore, to simulate the late-time evolution of the fallback rate for partial disruptions, we follow the same procedure as in \citet{Golightly:2019ab} and we replace the surviving core with a sink particle at a time significantly after the pericenter is reached. For simulations where there is a core that clearly survives the entire encounter intact, this time is $\sim 1.1$ days post-pericenter (for comparison, the sound-crossing time of our stars is $\lesssim 1$ hour). On the other hand, when $\beta \simeq \beta_{\rm c}$ and a small amount of mass remains in the core or the core reforms substantially after pericenter is reached, the time at which the core replacement occurs is significantly later (e.g., for the $1M_{\odot}$ ZAMS star with $\beta = 1.6$, we replace the core with a sink at a time of $\sim 134$ days post-pericenter). As also noted in \citet{Golightly:2019ab}, we have checked that changing the time at which the core is replaced does not alter the fallback rate in any noticeable way.

Finally the equation of state is isentropic and given by $P=K\rho^\gamma$, where $\rho$ is the density, $K$ scales with the entropy, and $\gamma$ is the adiabatic index. We take $\gamma=5/3$ and, following \cite{Golightly:2019ab}, the value of $K$ is a conserved quantity for each particle (as follows from the inviscid gas-energy equation) and is determined by the requirement of hydrostatic equilibrium for the density structure of the original star obtained from the {\sc mesa} outputs. We note that this means that the gas heats and cools under adiabatic expansion and contraction respectively, but no other forms of heating or cooling are included. For the weak encounters we present here shock heating of the gas is typically not important, and we have confirmed this by performing simulations in which shock heating is included and find no clear differences in the results. Following \cite{Coughlin:2015aa} we calculate the fallback rate by measuring the return rate of particles to the vicinity of the black hole. This is implemented by, after the initial disruption and the whole of the debris stream reaches a large radius, placing a sink radius around the black hole of size $\approx 3-5$ tidal radii with the value chosen such that it is sufficient to ensure that all returning material is captured by the sink radius.

\subsection{Results}
Our main aim in this work is to explore the properties of the fallback rates near the transition from full to partial disruption for different stars. We therefore start by identifying, for each simulated star, the largest $\beta$ for which a stellar core remains intact (or reforms as the debris recedes from the black hole) and the smallest $\beta$ for which the star is fully disrupted and remains so (i.e. there is no core present for the duration of the simulations).\footnote{In this respect we are referring to a single dominant core, and not the case where the debris stream locally fragments into many similar mass objects; see \citet{Coughlin:2015aa} and our discussion below.} For the $1M_\odot$ ZAMS star these are, respectively, {$\beta = 1.7$} and $\beta = 1.79$ (i.e., $\beta = 1.7$ has a core while $\beta = 1.79$ does not\footnote{The critical value of $\beta_{\rm c} \simeq 1.79$ for a $1M_{\odot}$, ZAMS star modeled with {\sc mesa} is consistent with the results of \cite{Law-Smith:2020aa} and is approximately 10\% smaller than the value found by \cite{Goicovic:2019aa}.}), for the $1M_\odot$ MAMS star these are {$\beta = 3.6$} and $\beta = 3.7$, and for the $0.3M_\odot$ MAMS star these are {$\beta=1.5$} and $\beta = 1.6$. The exact value of $\beta$ at which a core survives/reforms in the stream shows a small dependence on numerical resolution with simulations performed with $10^6$ particles. For example, we have simulated the initial disruption in a small number of cases with $10^7$ particles and found that the critical $\beta$ values vary by less than ten percent. Such small changes do not affect our results as here we are interested in the dynamical impact of the core on the stream dynamics; \cite{Miles:2020aa} show that $10^6$ particles is sufficient for this purpose (see their Figure 8). 

\begin{figure*}[t!]
\centering
	\includegraphics[width=\textwidth]{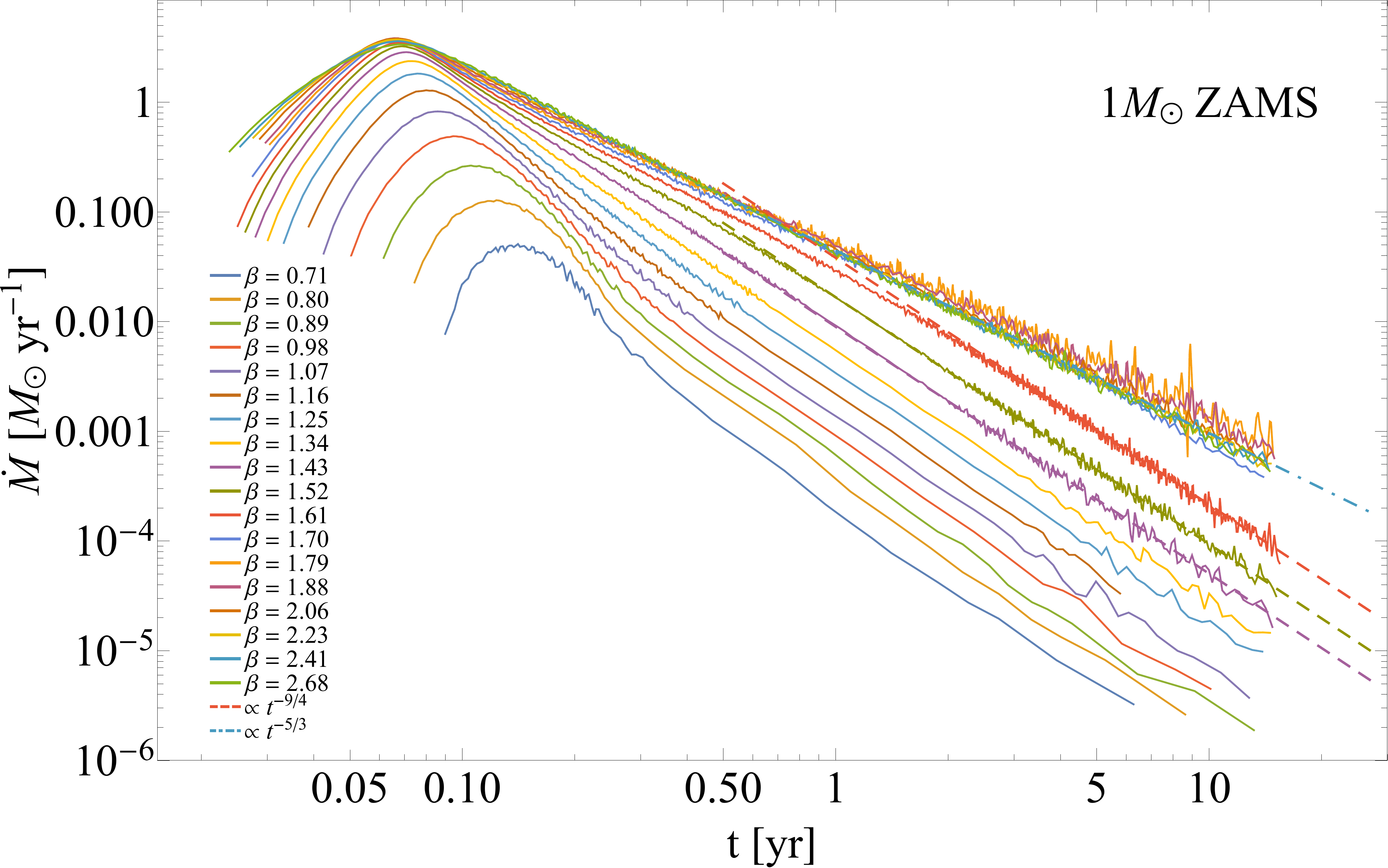}
		\caption{Fallback rates with time for the simulations with a $1M_\odot$ ZAMS star. The fallback rates are given in units of $M_\odot$ yr$^{-1}$ and time is in units of years. The line colours correspond to the $\beta$ values given in the legend. Lower values of $\beta$ generally correspond to lower accretion rates, particularly when $\beta < \beta_{\rm c}$, and thus typically the value of $\beta$ increases from the bottom curve to the top. Overlaid as a guide are the lines corresponding to $t^{-9/4}$ (dashed) and $t^{-5/3}$ (dot-dashed). For this star the value of $\beta_{\rm c}$ at which we find the star is just fully disrupted is $\beta_{\rm c} = 1.79$. There is a clear dichotomy between the cases where $\beta < \beta_{\rm c}$, which result in partial disruptions and a fallback rate that is consistent with $t^{-9/4}$, and the cases where $\beta > \beta_{\rm c}$, which result in full disruptions and a fallback rate that is consistent with $t^{-5/3}$.}
	\label{fig1}
\end{figure*}

\begin{figure*}[t!]
\centering
	\includegraphics[width=\textwidth]{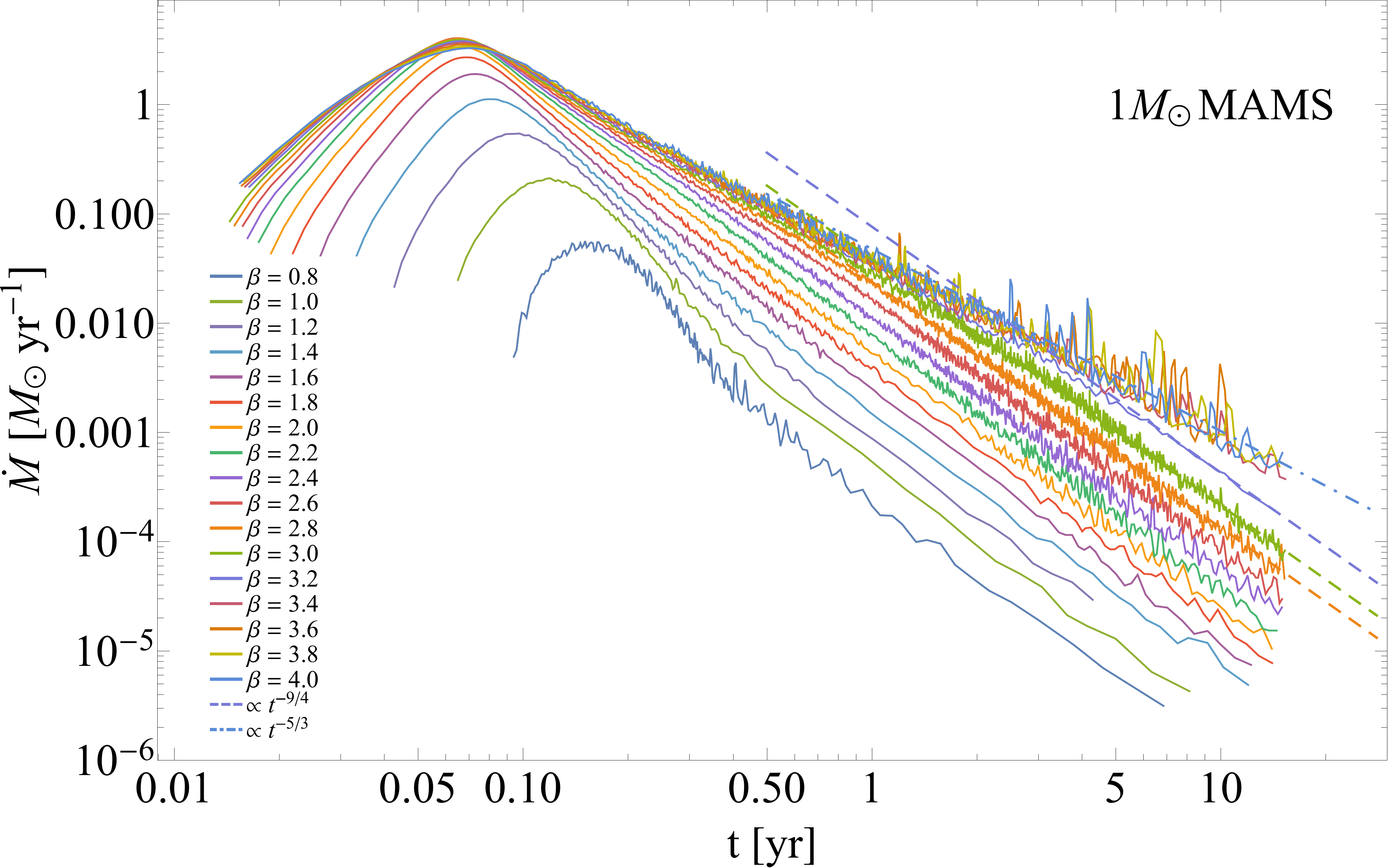}\vspace{0.05in}
	\includegraphics[width=\textwidth]{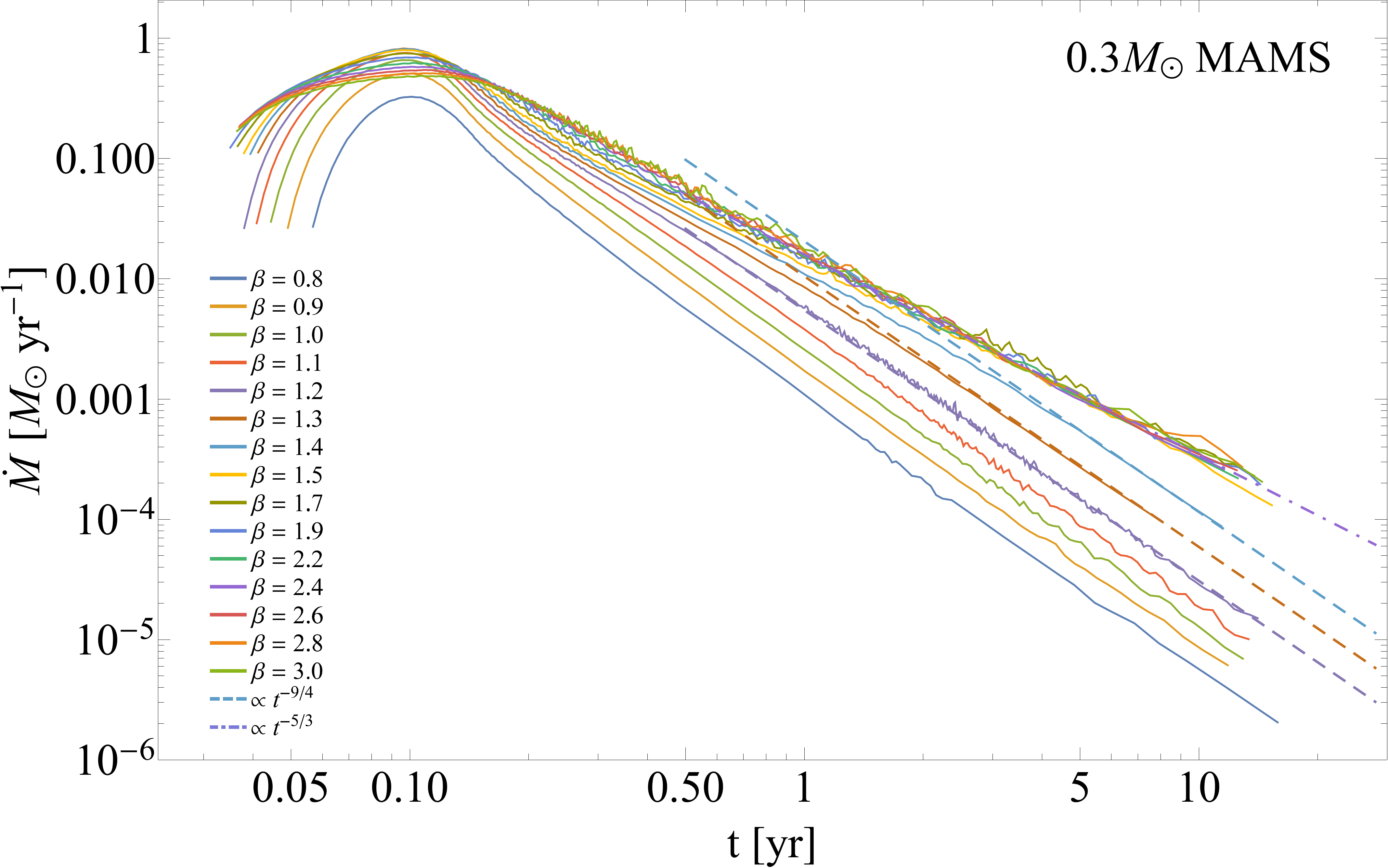}
	\caption{Same as Figure \ref{fig1} but for the $1M_{\odot}$ MAMS (top) and $0.3M_{\odot}$ MAMS (bottom) stars. For these stars $\beta_{\rm c} = 3.7$ (top) and $\beta_{\rm c} = 1.6$ (bottom).}
	\label{fig2}
\end{figure*}

In Figures \ref{fig1} and \ref{fig2} we present the fallback rates for a subset of the simulations for the $1M_{\odot}$ ZAMS star (Figure \ref{fig1}), the $1M_{\odot}$ MAMS star (Figure \ref{fig2}, top), and the $0.3M_{\odot}$ MAMS star (Figure \ref{fig2}, bottom). Only a subset is displayed so that the figures remain readable; the simulations not depicted show the same qualitative behaviour. In each figure the rate at which debris falls back to the black hole ${\dot M}_{\rm fb}$ is plotted in units of $M_\odot\,{\rm yr}^{-1}$ while time is measured in years. We compute the fallback rates explicitly from the simulations by measuring the rate at which particles return to the black hole\footnote{In particular we do not use the common approach of predicting forward from early times by using the assumption that the debris will continue on Keplerian orbits under only the black hole's gravity. For cases in which the stream contains a bound core, the assumption of Keplerian orbits with each fluid element possessing a conserved orbital energy is invalid \citep{Coughlin:2019aa}. Even when a core is not present there are times when this method yields inaccurate results, for example when the debris stream energy distribution is still evolving due to the effects of self-gravity \cite{Coughlin:2016ab,Coughlin:2016aa}.}. As the numerical method discretises the stellar debris into SPH particles, the fallback rate is subject to numerical noise on sufficiently small timescales. We therefore average the fallback rate in time following \cite{Miles:2020aa}, in that we bin the fallback rate in time at early times and by particle number at late times. We also employ an additional technique to bin over clumps when the stream fragments and produces physical (i.e., non-numerical) noise in the fallback curve; see Figure \ref{fig4} and the discussion thereof. 

\begin{figure}[h!] 
   \centering
   \includegraphics[width=\columnwidth]{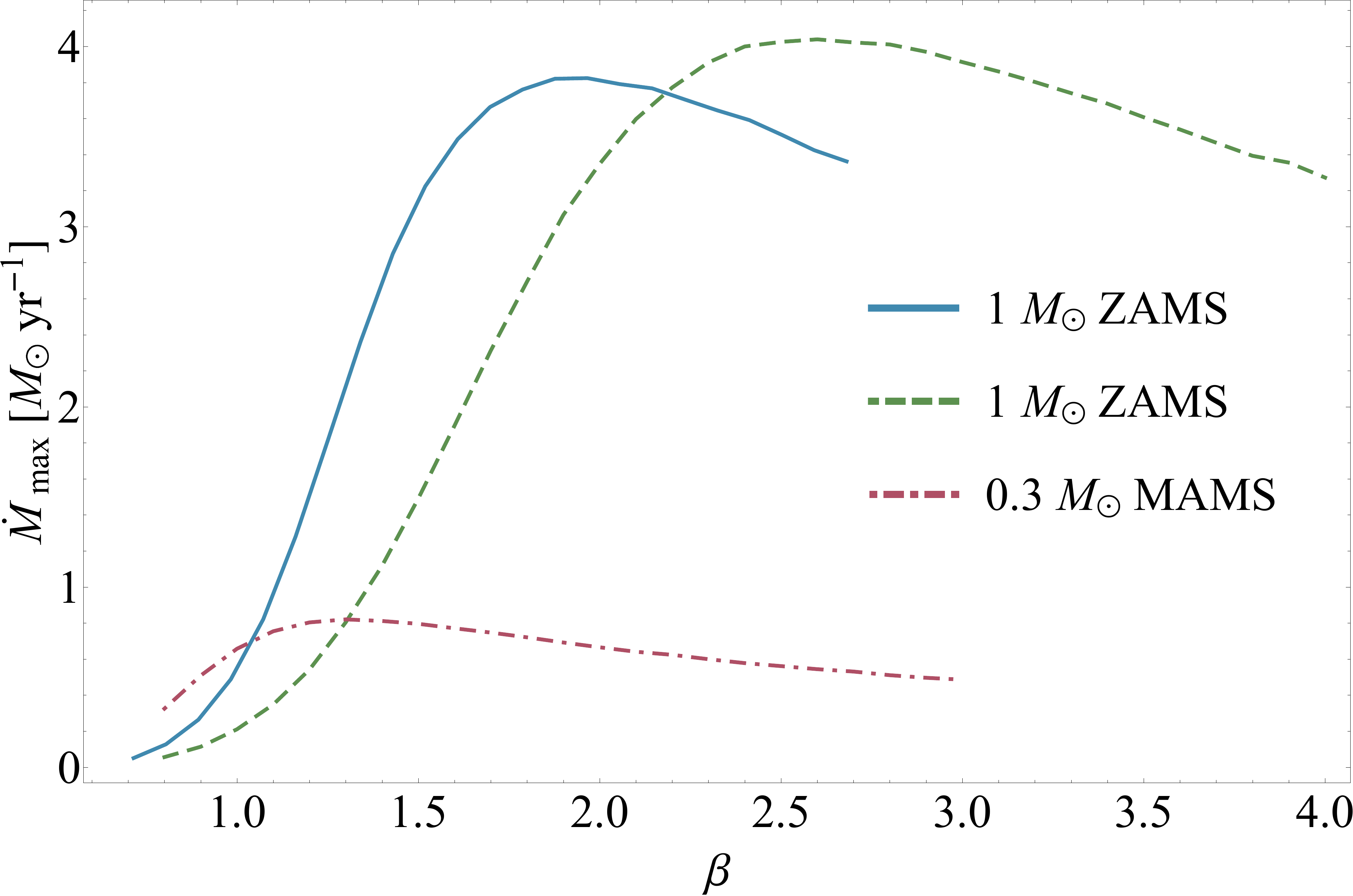} 
   \includegraphics[width=\columnwidth]{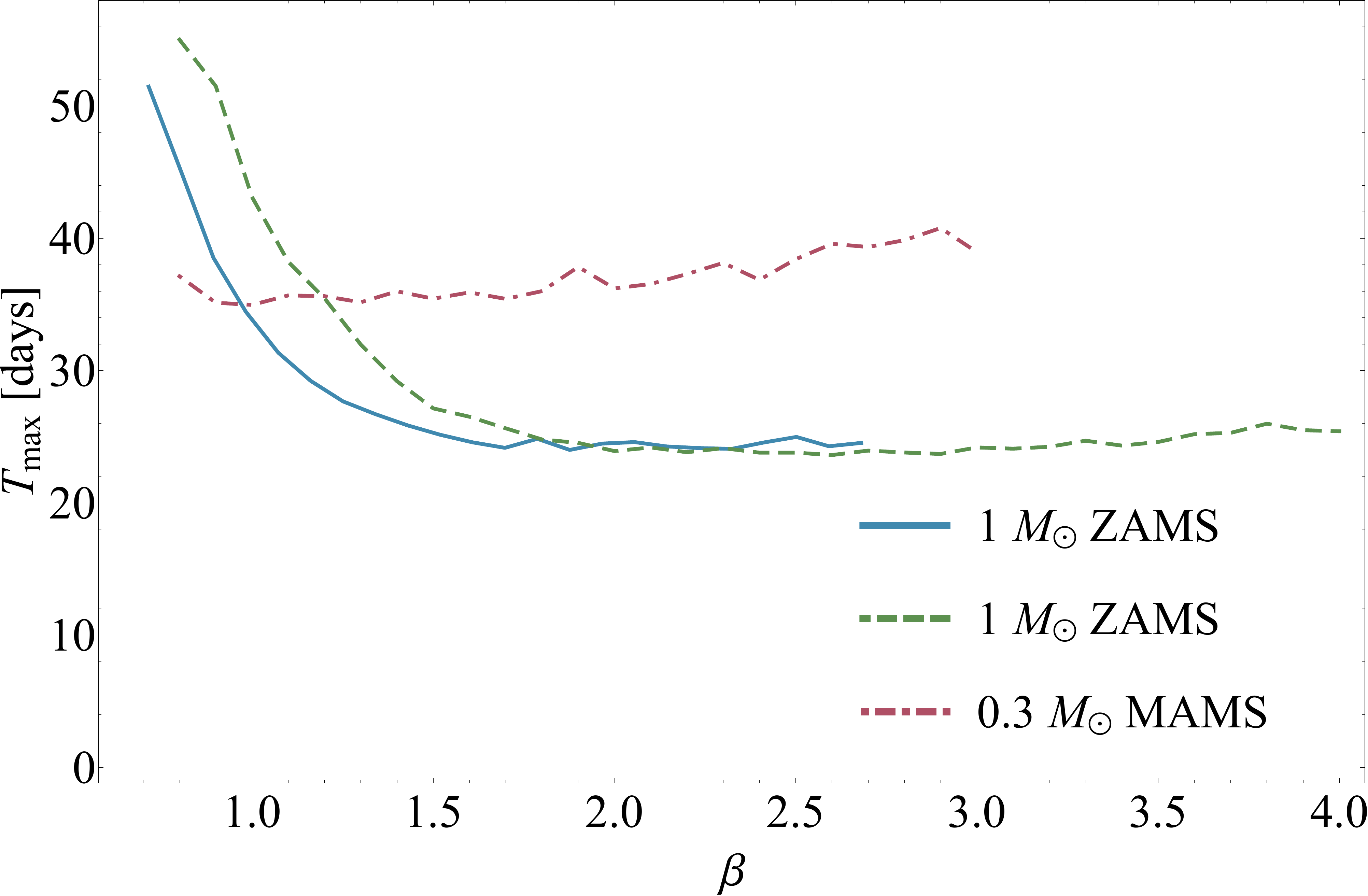} 
 \includegraphics[width=\columnwidth]{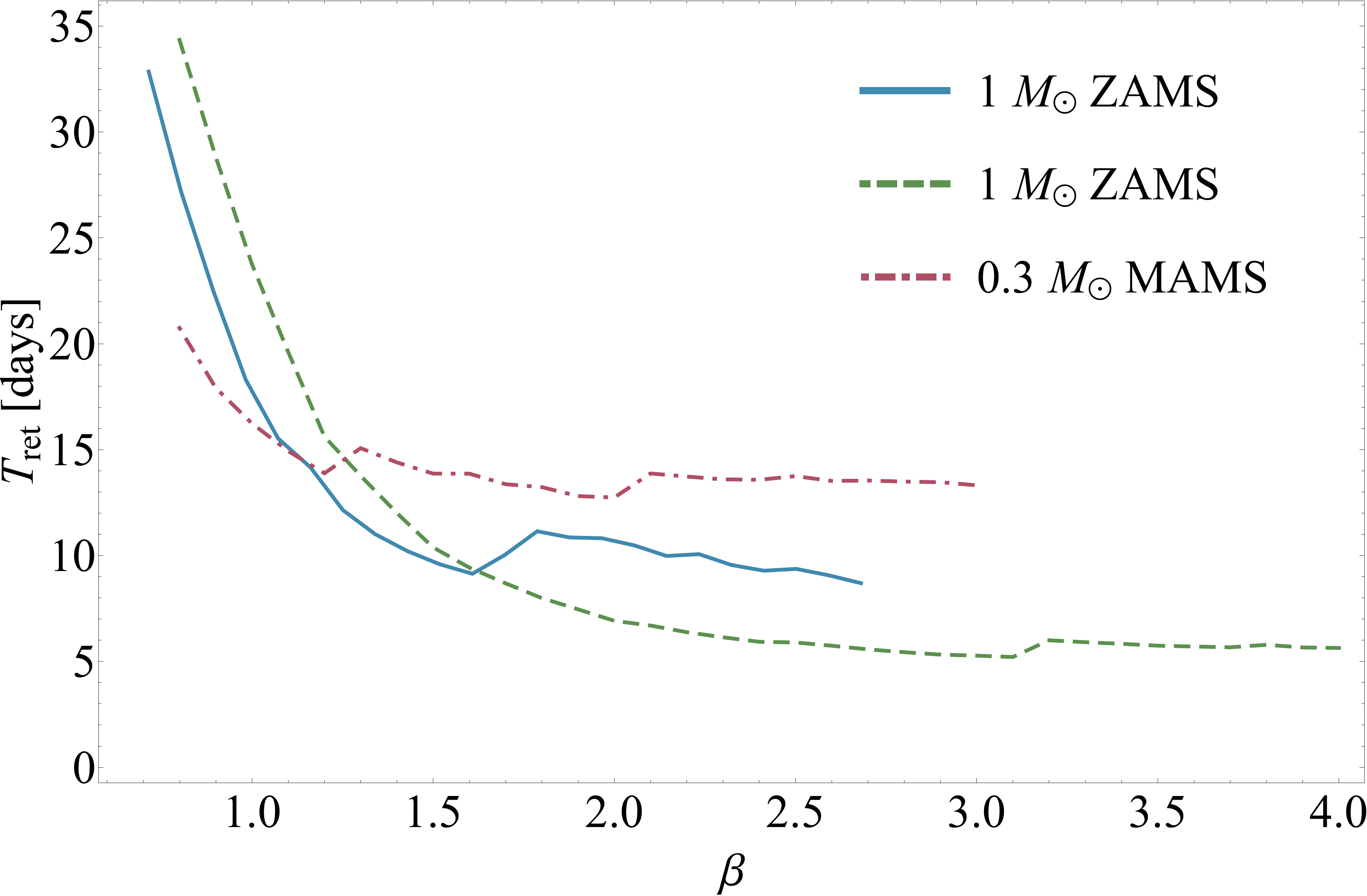} 
   \caption{The peak in the accretion rate (top), the time at which the peak in the accretion rate occurs (middle), and the return time of the most-bound debris (bottom) for the 1$M_{\odot}$ ZAMS star (blue, solid), the $1M_{\odot}$ MAMS star (green, dashed), and the $0.3M_{\odot}$ MAMS star (red, dot-dashed) as functions of $\beta$.}
   \label{fig3}
\end{figure}

\begin{figure}
	\includegraphics[width=\columnwidth]{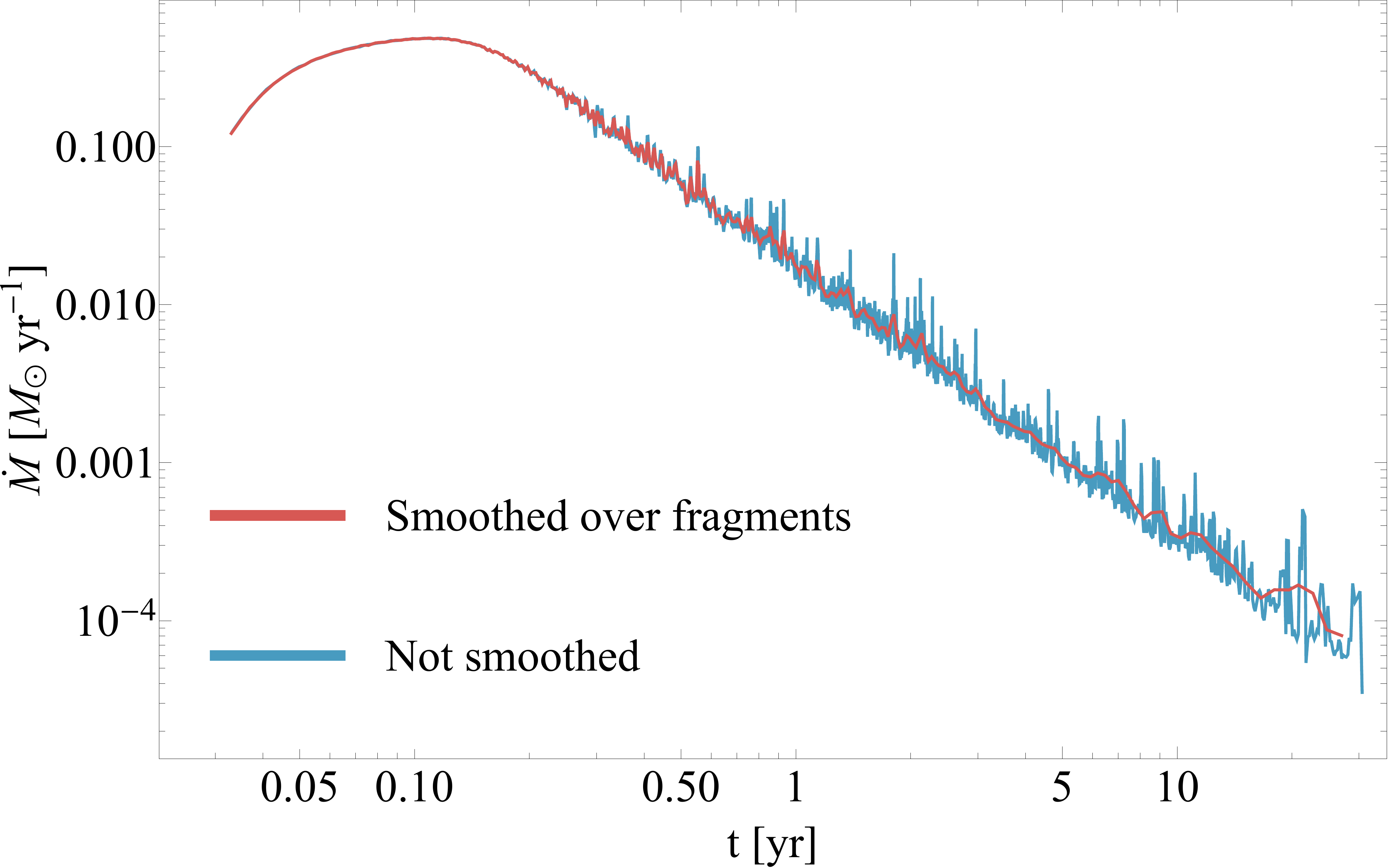}
	\caption{The fallback rate for the $\beta = 3.0$ disruption of the 0.3 $M_{\odot}$, MAMS star. The red line depicts the smoothed fallback rate (as shown in the bottom panel of Figure \ref{fig2}), while the blue line shows the un-smoothed fallback rate. Similar to the fallback rate shown presented in \cite{Coughlin:2015aa}, we see that there are spikes at times $\gtrsim 0.5$ years, which are associated with the accretion of clumps of gas that have formed within the debris stream. This indicates that the stream is gravitationally unstable. In this case this has occurred for a stream produced from a TDE with $\beta = 2\beta_{\rm c}$. }
	\label{fig4}
\end{figure}

As one might expect, the accretion rates are generally lower for less disruptive encounters (smaller $\beta$) in which the mass of the debris stream is a smaller fraction of the mass of the original star. For full disruptions we generally find that the time at which the fallback begins and the time and magnitude of the peak fallback is very similar over a factor of $\approx 2$ in impact parameter, indicating that the width of the spread of orbits occupied by the debris is not strongly dependent on $\beta$ \citep[see, e.g.,][]{Stone:2013aa}; this behavior was also found by \citet{Guillochon:2013aa}. For partial disruptions of the $1M_{\odot}$ stars (both ZAMS and MAMS) we find a noticeable delay in the onset of fallback, which occurs progressively later and with a lower peak rate as the disruption is made weaker (decreasing $\beta$). For the $0.3M_\odot$ MAMS star the time at which the fallback rate peaks does not vary significantly across the entire $\beta$ range we simulated. To substantiate these notions, the top, middle, and bottom panels in Figure \ref{fig3} respectively show the peak in the accretion rate $\dot{M}_{\rm max}$, the time at which the peak occurs $T_{\rm max}$, and the return time of the most-bound debris $T_{\rm ret}$ for all three stellar progenitors over the range of $\beta$ that we simulated for each star. The feature of Figures \ref{fig1} and \ref{fig2} that we emphasize most is that for all of the simulations the fallback rates approach a clear power-law decay, and this power-law is well-described by one of two modes: either $n \approx -5/3$ for full disruptions or $n\approx -9/4$ for partial disruptions. To substantiate this result further, we present fits to the fallback data and plots of the power-law index with time, $n(t)$, in Section~\ref{fits}.

In addition to this main result, we also find several other interesting features of the simulations, the first being that self-gravity can remain important in determining the structure of the debris stream after a star has undergone a full disruption, and that in some cases this can lead to widespread fragmentation of the debris stream. This result was first presented by \cite{Coughlin:2015aa}, who found that the debris stream could fragment under its own self-gravity for a full disruption of a solar-like star modelled as a $\gamma=5/3$ polytrope on a parabolic orbit with $\beta=1$. The subsequent dynamics of self-gravitating debris streams has been explored by \cite{Coughlin:2016ab,Coughlin:2016aa}, and the gravitational instability of hydrostatic filaments that is fundamentally responsible for the fragmentation is detailed in \cite{Coughlin:2020aa}. More recently, \cite{Coughlin:2020ac} presented a semi-analytical model for the evolution of stellar debris orbits under the assumption that they evolve ballistically following the disruption of the star. Using this model they showed that the debris evolves through an in-plane caustic \citep[discovered by][]{Coughlin:2016aa}, that is capable of augmenting the density of the debris back above the tidal density for {\it any} value of $\beta$; the continued importance of self-gravity -- even for relatively high-$\beta$ disruptions -- was also found by \citet{Steinberg:2019aa}, who simulated the disruption of $\gamma = 4/3$ and $\gamma = 5/3$ polytropes up to $\beta = 7$. We show in Figure \ref{fig4} the fallback rate for the disruption of the $0.3M_{\odot}$ MAMS star with $\beta = 3.0$, and therefore for which $\beta/\beta_{\rm c} \simeq 2$. As described above, to make the fallback rates in Figures \ref{fig1} and \ref{fig2} we smoothed the data over the fallback of clumps that form out of the fragmentation of the tidally disrupted debris stream. In this Figure \ref{fig4} we show both the clump-smoothed and un-smoothed fallback data. In the un-smoothed case the accretion of clumps of gas within the debris stream are clearly visible at times $t\gtrsim 0.5$ years. This shows that the stream is self-gravitating and vulnerable to fragmentation for moderate $\beta$ values with $\beta = 2\beta_{\rm c}$. 

\begin{figure*}[htbp] 
   \centering
   \includegraphics[width=0.495\textwidth]{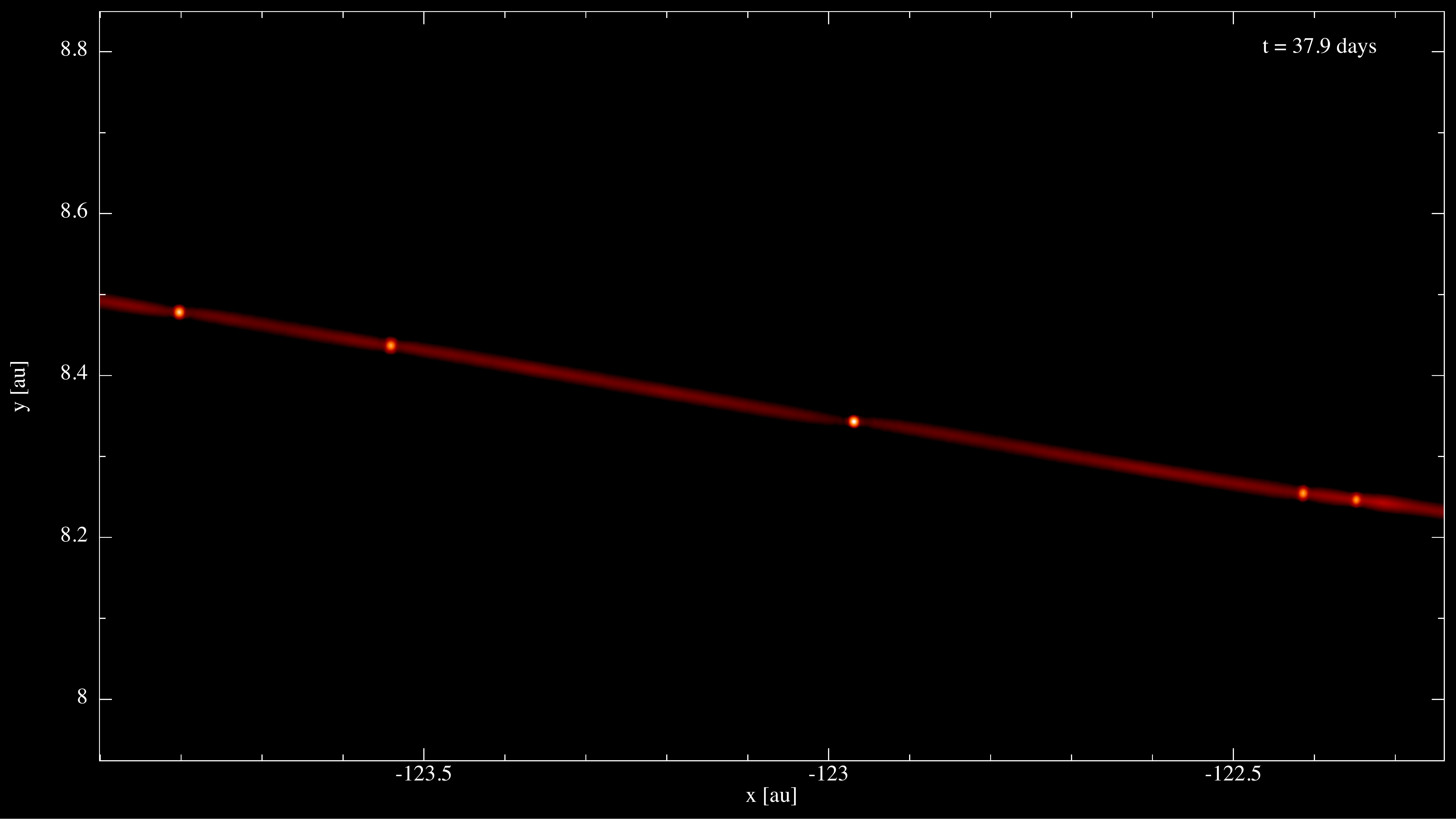} 
   \includegraphics[width=0.495\textwidth]{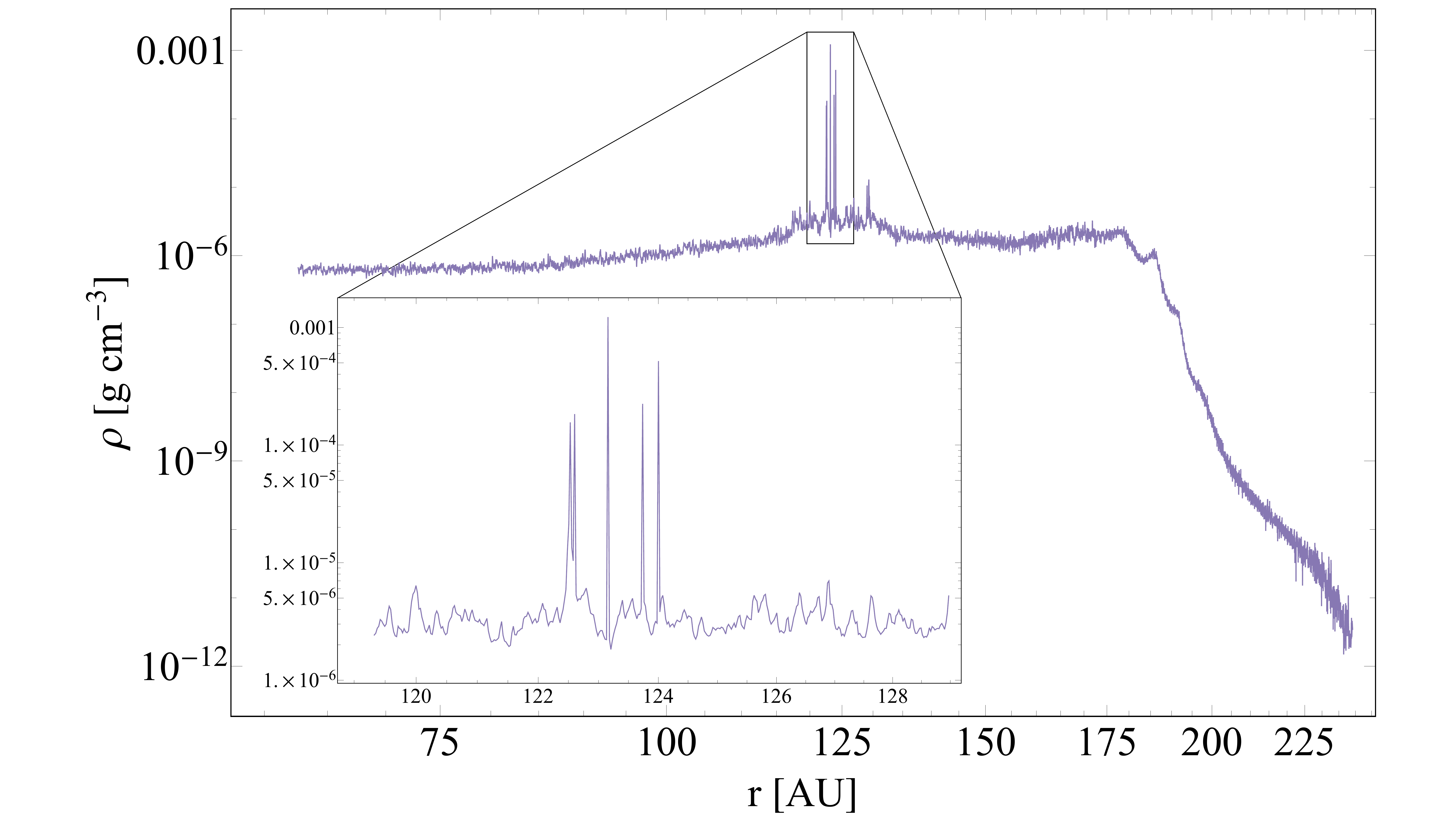} 
   \caption{Left: A rendering of the density of the stream formed out of the $0.3 M_{\odot}$, MAMS, $\beta = \beta_{\rm c} = 1.6$ disruption, with brighter colors indicating regions of enhanced density. This figure illustrates that, instead of recollapsing to a single core (which occurs for $\beta = 1.4$ and $\beta = 1.5$), the stream fragments into five cores that are highly clustered near the maximum in the density of the original stream. Right: The density along the stream as a function of distance from the supermassive black hole (the density is averaged over the small solid angle subtended by the stream), with the inset giving a close-up view of the five, highly localized maxima in the density that correspond to the locations of the five cores that collapse out of the stream.}
   \label{fig5}
\end{figure*}

\begin{figure*}[htbp] 
   \centering
   \includegraphics[width=0.495\textwidth]{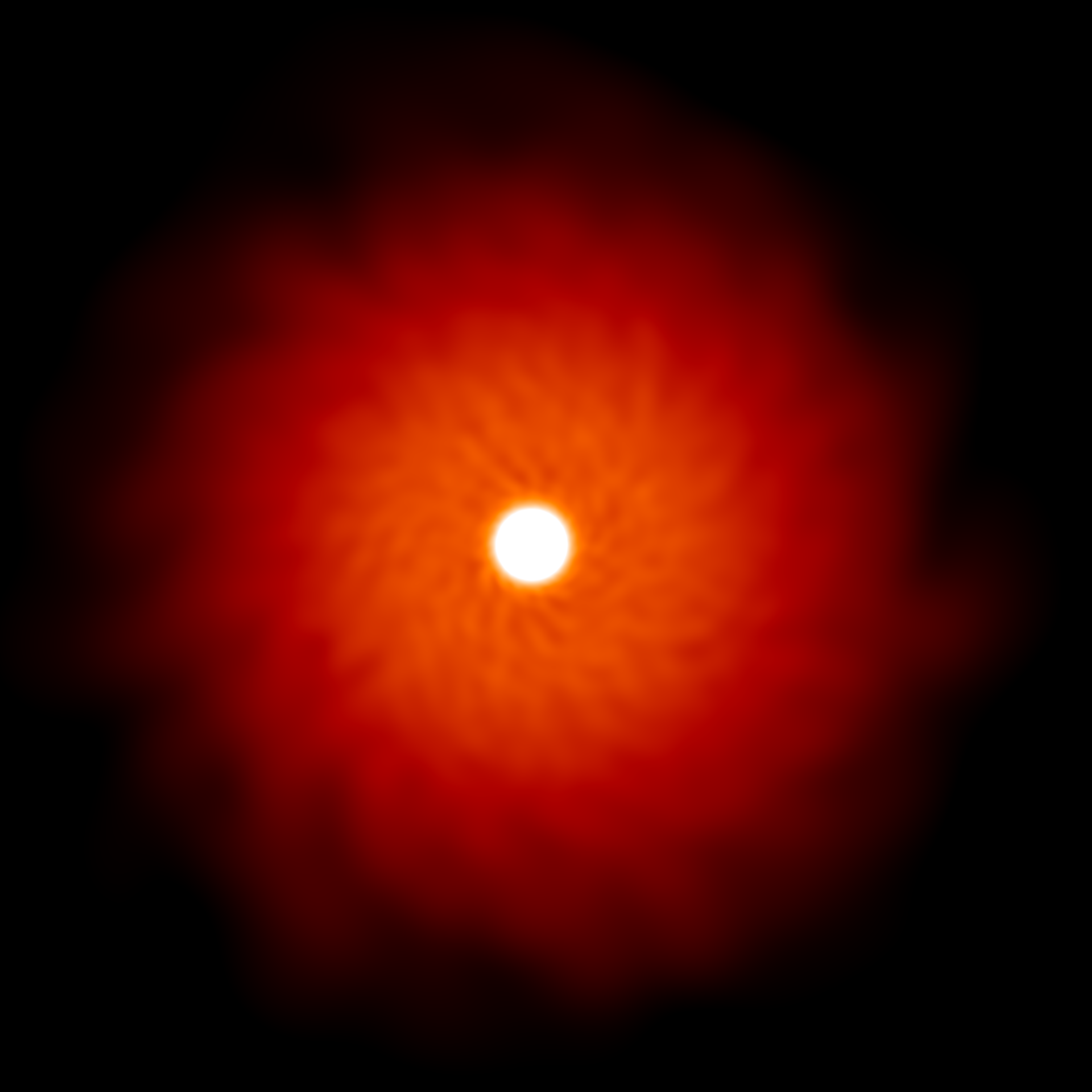} 
   \includegraphics[width=0.495\textwidth]{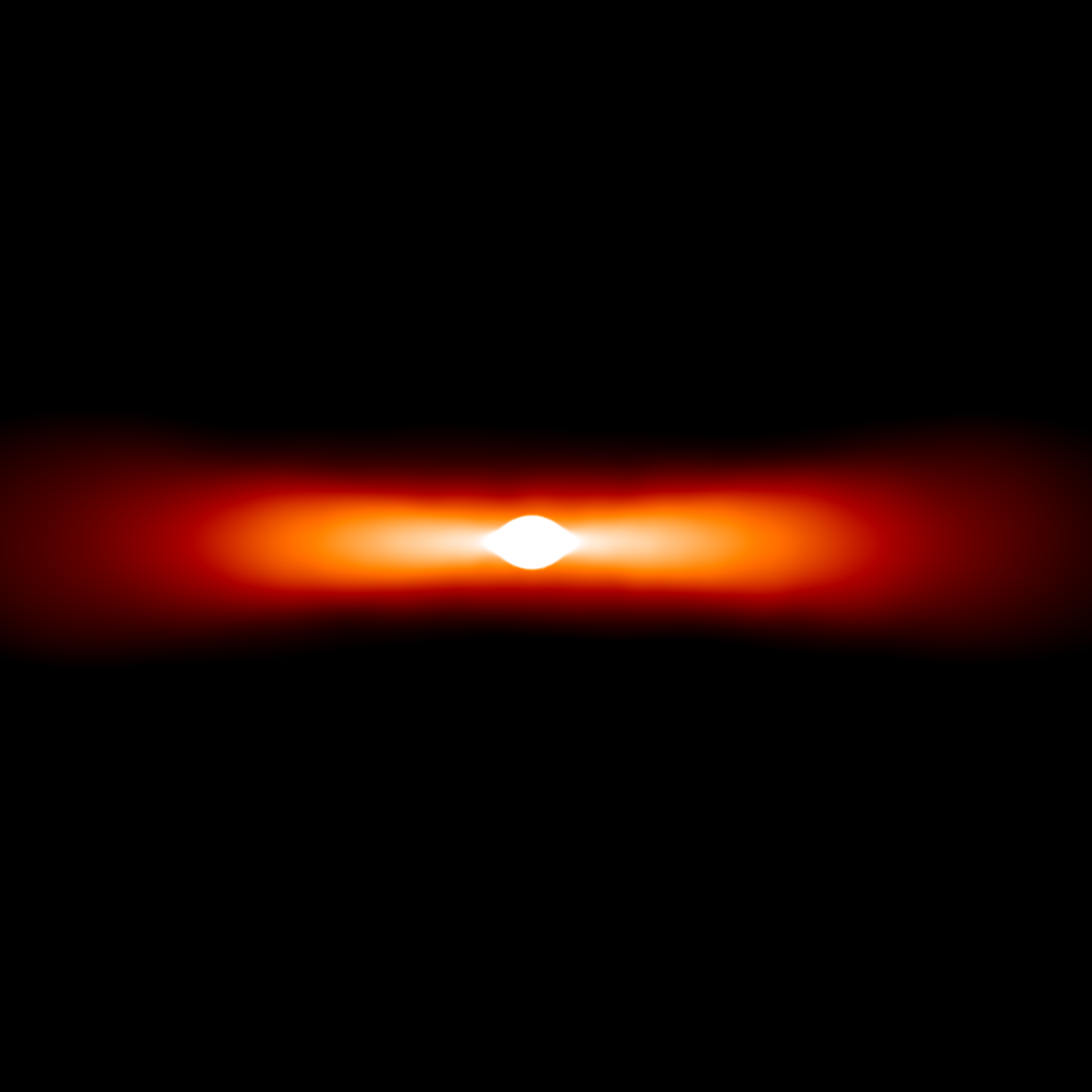} 
   \caption{A rendering of the disc formed around the surviving stellar core of the $\beta = 1.61$ disruption of the $1M_{\odot}$, ZAMS star. The left image gives the projection of the density onto the orbital plane of the original star, with brighter (darker) regions indicating areas of enhanced (reduced) density, while the image on the right gives the projection out of the plane.}
   \label{fig6}
\end{figure*}

\begin{figure*}[htbp] 
   \centering
   \includegraphics[width=0.495\textwidth]{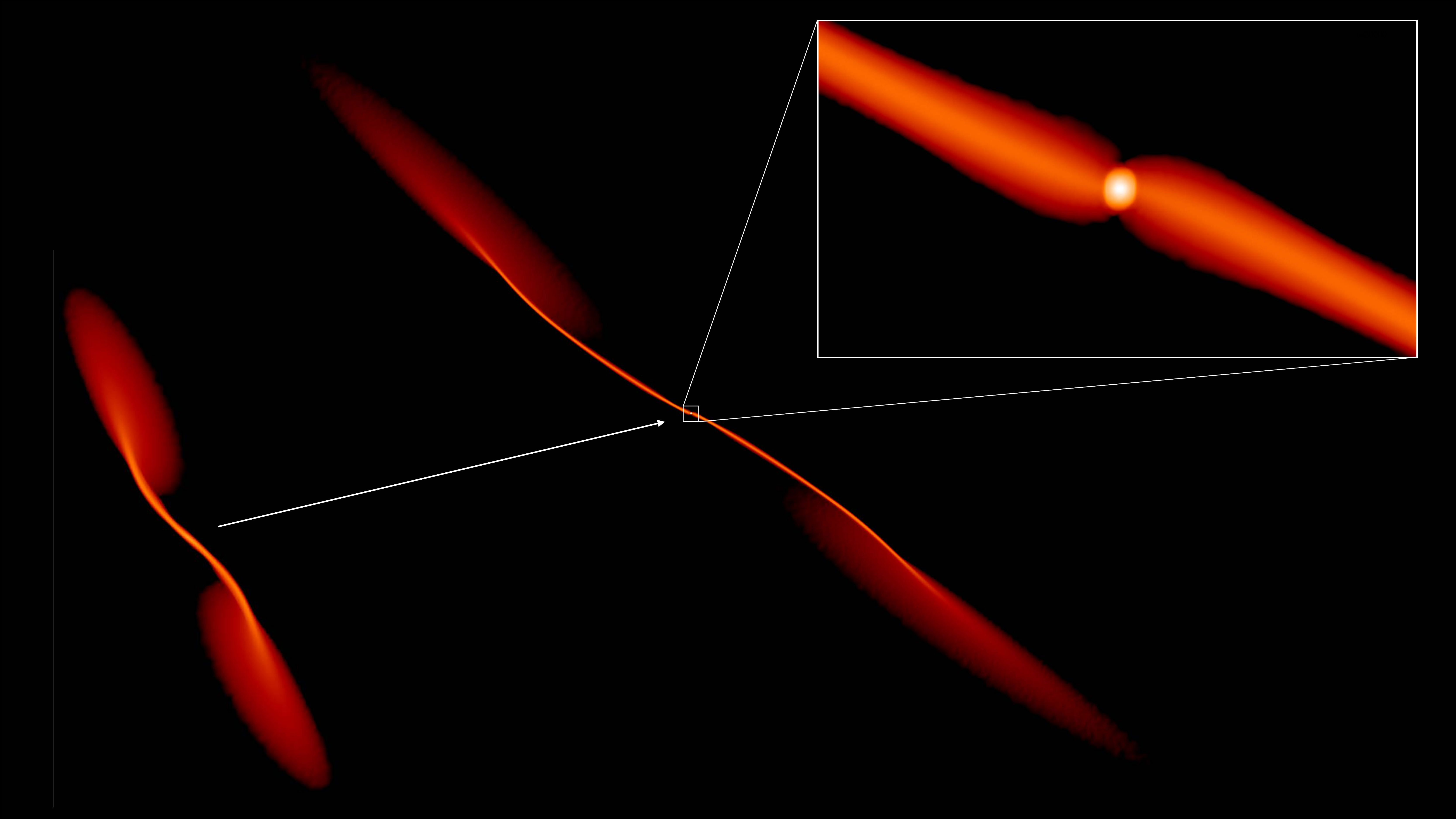} 
   \includegraphics[width=0.495\textwidth]{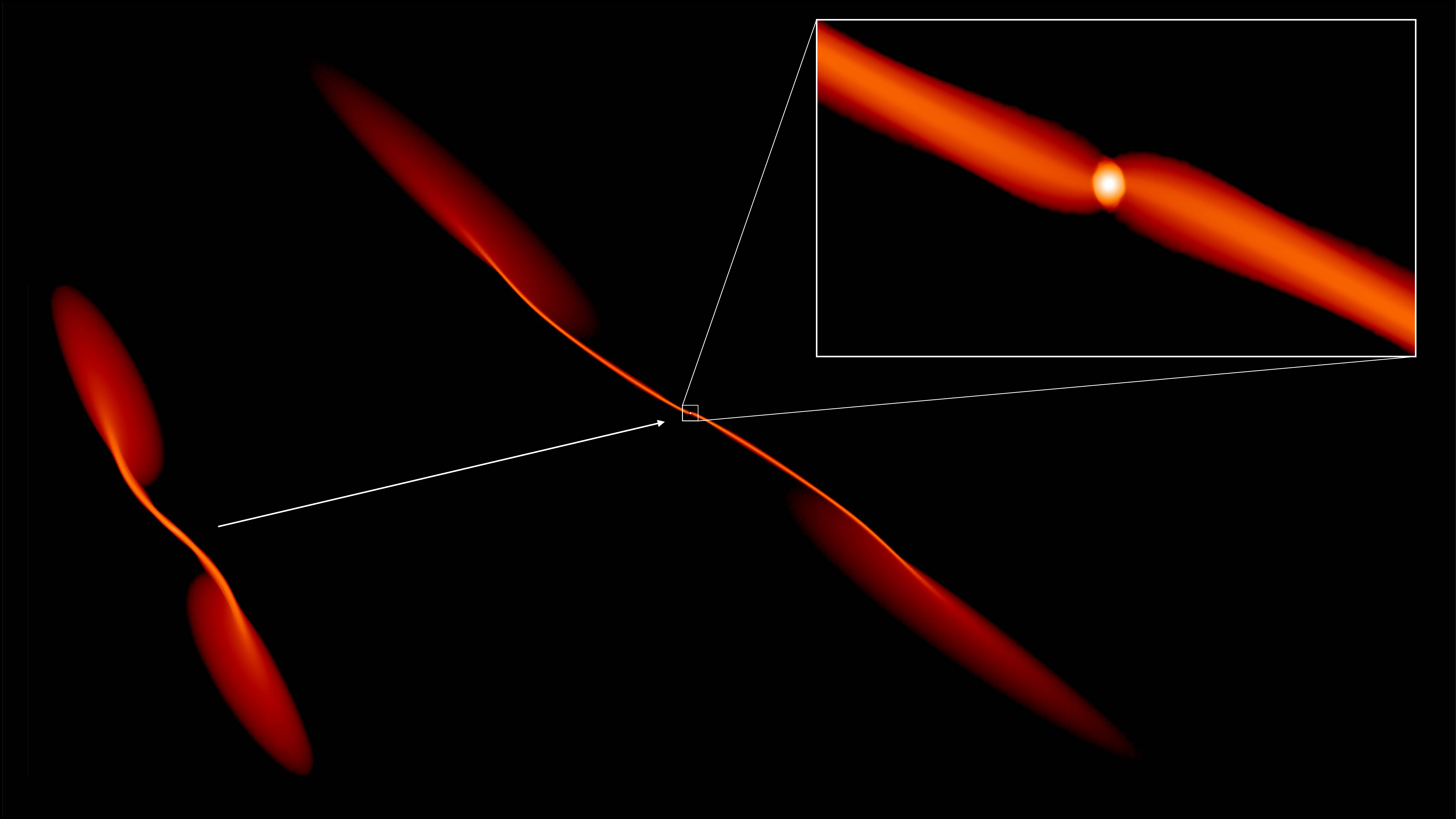} 
   \caption{The $\beta = 1.5$ disruption of the $0.3 M_{\odot}$ star in which a core reforms after the star is initially completely disrupted, where the left panel was performed with $10^6$ particles (the resolution we used for the results in this paper) and the right panel with $10^7$ particles. The stream in the bottom-left corner of each panel is at a time of $\sim 0.29$ days post-pericenter, while the stream in the middle of each panel (which has a reformed core, shown by the inset in the top-right of each panel) is at a time of $\sim 1.1$ days post-disruption. We see excellent agreement between the two resolutions. (A higher-resolution version of this image can be obtained from the authors on request.)}
   \label{fig:zombie}
\end{figure*}

Self-gravity also, and not surprisingly, plays a crucial role in the determination of the final state of the debris stream when $\beta \simeq \beta_{\rm c}$. For disruptions that are just below $\beta_{\rm c}$, we find that the star is initially completely disrupted by the black hole, with no core surviving the encounter intact by $\sim 1.1$ days post-disruption (i.e., the time at which the core would be replaced by the sink to study the long-term behavior of the fallback rate). At a time significantly later than this, however, a {zombie} core reforms out of the disrupted debris; this behavior was also seen in \citet{Guillochon:2013aa}. Figure \ref{fig:zombie} illustrates the $\beta = 1.5$ disruption of the $0.3M_{\odot}$ star, one such case in which this recollapse is observed, where the simulation in the left panel was performed with $10^6$ particles -- being our standard resolution and the one used for the simulation results presented in this paper -- while the one in the right panel uses $10^7$ particles. The stream in the bottom-left of each panel, which clearly does not have a stellar core, is at a time of $\sim 0.29$ days post-disruption. The stream in the center of each panel, which has a core near its geometric center (as shown by the inset in the top-right of each panel), is at a time of $\sim 1.1$ days post-disruption. In addition to providing a specific example of a TDE in which this recollapse occurs, this figure also demonstrates that our simulations at $10^6$ particles accurately capture the physics of the disruption and core reformation.

For disruptions at $\beta_{\rm c}$ and slightly larger, a core tries to reform, but instead the stream's self-gravity causes it to fragment on a smaller scale (see the discussion in \citet{Coughlin:2020aa} and \citet{Coughlin:2020ab} about the wavenumber at which the instability growth rate peaks in magnitude and how this consideration applies to short gamma-ray bursts, respectively). Interestingly, for the $0.3M_{\odot}$, MAMS star with $\beta = \beta_{\rm c} = 1.6$, we find that instead of recollapsing to a single core, the stream fragments into \emph{five} massive cores that are all localized near the maximum in the stream density. The left panel of Figure \ref{fig5} shows a density rendering of the stream at $\sim 1$ month post-disruption, with brighter (darker) regions indicating areas of larger (smaller) density. The right panel shows the density (in g cm$^{-3}$) as a function of distance (in AU) from the supermassive black hole, where here we averaged over the small solid angle subtended by the stream to obtain a density purely as a function of $r$. Both of these images serve to illustrate the impact of self-gravity on the stream, and provide substantial evidence to suggest that the same instability (cf.~\citealt{Coughlin:2020aa}) is responsible for the recollapse of the stream into a single core and the fragmentation of the debris stream into many, small-scale knots.

We also find that the surviving core in some simulations can be surrounded by and maintain a circumstellar disc of material.\footnote{Previously \cite{Sacchi:2019aa} have shown that a TDE involving an initially rapidly rotating star whose rotation axis is retrograde to the orbital axis around the black hole can result in the formation of a circumstellar disc. Here we show that disc formation may be possible for initially non-rotating stars, and thus may be a much more likely result from typical TDEs. In subsequent work, we will explore the likelihood of disc formation and the implications for observability for stars in our galaxy.} Figure \ref{fig6} shows one such disc from the $1M_{\odot}$, ZAMS, $\beta = 1.61$ disruption, where the left (right) panel is the projection of the density on to (out of) the orbital plane of the original star; as for the left panel of Figure \ref{fig5}, denser regions in this figure are more brightly colored. In this simulation the core does not survive the initial encounter intact, and instead reforms out of the stream at a time of roughly $\sim 0.5$ days post-disruption. By following the temporal evolution of the core and the disc, we find that most of the disc is composed of material that was initially in the very outer layers of the reformed core, though a smaller fraction of the disc mass is supplied by material that continues to rain down on the core (from the tidally shed debris stream) at later times.  This finding demonstrates that this disc of material is ``spun off'' the surviving core. This finding is reminiscent of the discs formed around rapidly rotating B-type stars (Be-stars; see the review by \citealt{Rivinius:2013aa}).\footnote{Although we note that it is generally accepted that the discs in Be stars are not formed in this way as the rotation rates are typically not high enough \citep{Rivinius:2013aa}. Be star discs may instead form due to the input of mass from small-scale magnetic flaring events on the stellar surface. These, combined with the observed rapid rotation, can provide the mass and angular momentum required to form and sustain a Keplerian Be-star disc \citep{Nixon:2020aa}.}

\section{Analytical fits to the fallback data and \lowercase{$n(t)$} curves}
\label{fits}
The main result presented above is that for stars modelled with accurate density profiles and with a finely sampled $\beta$ parameter there is a clear dichotomy between the late-time power-law indices of the fallback rates, with those simulations in which $\beta > \beta_{\rm c}$ (i.e. full disruptions) exhibiting a $\propto t^{-5/3}$ decay and those with $\beta < \beta_{\rm c}$ (i.e. partial disruptions) exhibiting a $\propto t^{-9/4}$ decay. To substantiate this further, we provide here analytical fits to the fallback data, and the resulting evolution of the power-law index with time, $n(t) = d\log(\dot{M})/d\log(t)$. With these fits we recover the dichotomy shown in Figures \ref{fig1} and \ref{fig2} and are able to confirm that the late-time power-laws $n(t\rightarrow\infty)$ are generally consistent with either $-5/3$ or $-9/4$ as predicted by \cite{Coughlin:2019aa}. There are a small number of simulations in which there is a small, but significant, offset from these powerlaws, and these cases---which occur when $\beta \simeq \beta_{\rm c}$---appear to be those in which self-gravity is still acting to re-arrange the mass distribution along the debris stream by the end of our simulations. Here we provide the methodology used to fit the fallback rates and the resulting fits.

We fit the fallback curves to the following, modified Pad\'e approximant:
\begin{equation}
\dot{M}_{\rm fit} = \frac{a\tilde{t}^{m}}{1+\frac{a}{b}\tilde{t}^{m-n_{\infty}}}\frac{1+\sum_{i = 1}^{N_{\rm max}-1}c_{\rm i}\tilde{t}^{i}+\tilde{t}^{N_{\rm max}}}{1+\tilde{t}^{N_{\rm max}}}. \label{Mdotfit}
\end{equation}
Here $\tilde{t} = t/t_{\rm max}$, where $t_{\rm max}$ is the time at which the fallback rate reaches a peak, and $N_{\rm max} \ge 1$ (when $N_{\rm max} = 1$ there are zero terms in the sum by definition). The motivation for this functional form is the following: at times earlier than the peak in the fallback rate we expect $\dot{M}$ to rise as a power-law in time $\propto t^{m}$, while at times much greater than the peak it will fall off as a power-law $\propto t^{n_{\infty}}$. These considerations suggest that the early ($\tilde{t} \ll 1$) and late-time ($\tilde{t} \gg 1$) behavior can be well-fit by the function $a\tilde{t}^{m}/(1+a/b\times\tilde{t}^{m-n_{\infty}})$. In between the initial rise and eventual decay the fallback rate peaks and shows additional variation; these additional variations can be captured by the ratio of polynomials in $\tilde{t}$ in Equation \eqref{Mdotfit}, and increasing the number of terms in the sum in the numerator (increasing $N_{\rm max}$) leads to more of these features being accurately reproduced by the function $\dot{M}_{\rm fit}$. The ratio of polynomials is the Pad\'e approximant.

The parameters $a$, $m$, $b$, $n_{\infty}$, $c_1$, $c_2$, \ldots, $c_{\rm N_{\rm max}-1}$ are fit by minimizing the $\chi^2$ of the logarithm of the data, i.e., the parameters minimize
\begin{equation}
\chi^2 = \sum_i\left(\ln\left(\frac{\dot{M}_{\rm i}}{\dot{M}_{\rm half max}}\right)-\ln\left(\frac{\dot{M}_{\rm fit}(\tilde{t}_{\rm i})}{\dot{M}_{\rm halfmax}}\right)\right)^2.
\end{equation}
In this sum $\dot{M}_{\rm i}$ is the numerically obtained fallback rate at the time $\tilde{t}_{\rm i}$. We emphasize the importance of minimizing the $\chi^2$ of the \emph{log} of the data as opposed to the data itself: the latter does not accurately constrain the late-time falloff of the fallback rate as $\dot{M}_{\rm i} \ll \dot{M}_{\rm max}$ at these times. Therefore, the fitted function could be anything $\ll \dot{M}_{\rm max}$ at these late times and the contribution to the $\chi^2$ of the non-logged data would still be small. By minimizing the $\chi^2$ of the logarithm of the data, on the other hand, the late-time behavior retains its importance in terms of its contribution to the $\chi^2$. We also normalize the fallback rate by its half-max value ($\dot{M}_{\rm halfmax} = \dot{M}_{\rm max}/2$) so that both the peak data and the late-time data are comparable in magnitude.

\begin{figure*}
	\includegraphics[width=0.495\textwidth]{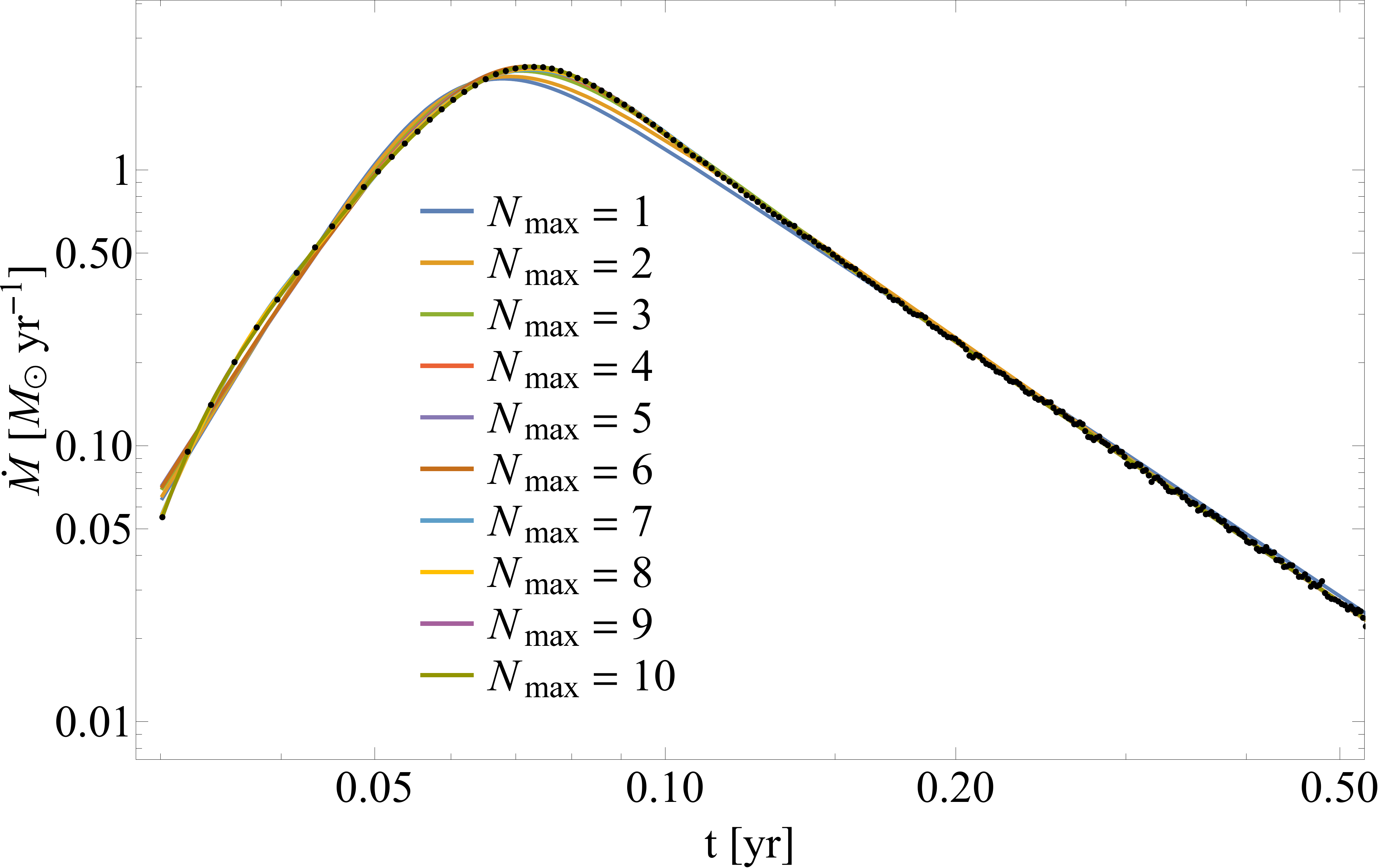}
	\includegraphics[width=0.495\textwidth]{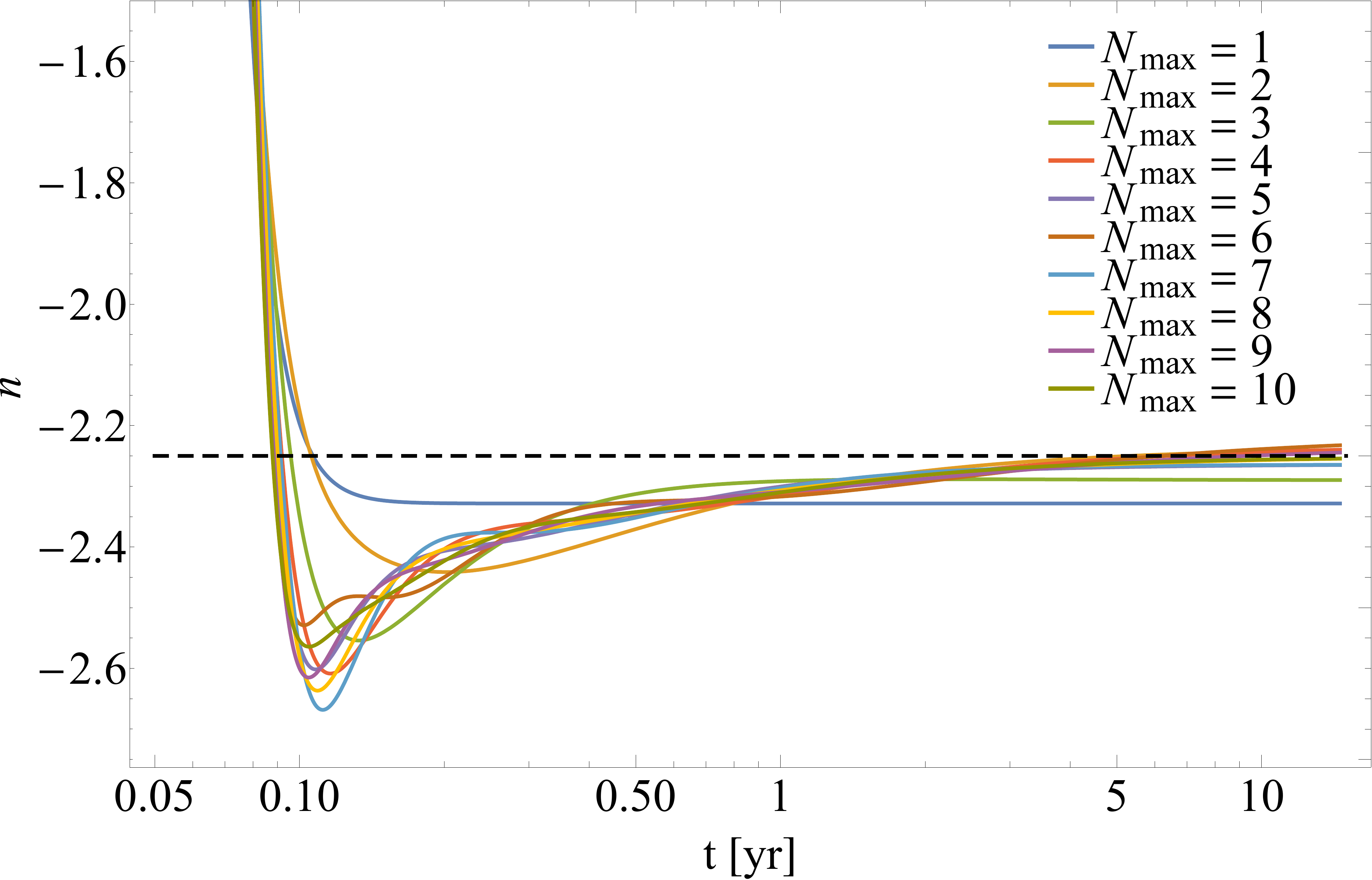}
	\caption{Comparison of the fallback rate $\dot{M}$ (left) and the instantaneous power-law index of the fallback rate $d\ln\dot{M}/d\ln t \equiv n(t)$ (right) for different terms in the series expansion of the fitting function (Equation \ref{Mdotfit}) for the $1M_{\odot}$ ZAMS star with $\beta = 1.34$. The black points in the left panel show the fallback data for this simulation, while the dashed line in the right panel shows $n = -9/4$.}
	\label{fig7}
\end{figure*}

The left panel of Figure \ref{fig7} shows the fallback rate from the $1M_{\odot}$, ZAMS, $\beta = 1.34$ simulation (points) and fits with varying $N_{\rm max}$ (curves; the legend gives $N_{\rm max}$); here we zoomed in on the near-peak behavior of the fallback as this is where there is the most visible variation in the curves. As we increase the number of terms in the expansion the fit to the data gets noticeably better, but it is apparent that once $N_{\rm max} \gtrsim 5$ the degree to which the fit follows the data does not increase appreciably. The latter statement is quantified by the penultimate column of Table \ref{tab:1}, which gives the $\chi^2$ of the log of the data; we see that in going from 0 to 1 terms in the expansion ($N_{\rm max} = 1$ to 2), 1 to 2 terms, and 2 to 3 terms, the relative decrease in $\chi^2$ is (respectively) $\Delta\chi^2/\chi^2 \simeq -48.1\%$, $-26.5$\%, and $-8.21\%$. Therefore, as we add more terms to the series expansion the $\chi^2$ is reduced, but the degree to which it is reduced gradually lessens as we increase $N_{\rm max}$. The right panel of this figure shows the instantaneous power-law index of the fallback rate $d\ln\dot{M}/d\ln t \equiv n(t)$. From this figure the difference in the solutions as we increase the number of terms is more apparent, with more nuanced behavior of the derivative of the fallback rate being better-captured by solutions with more terms.

By construction the asymptotic power-law index of the fallback rate is given by $n_{\infty}$ in the fitting function in Equation \eqref{Mdotfit}. The uncertainty of $n_{\infty}$ is defined as\footnote{We also implemented a second technique in which we ``jackknifed'' the data -- we randomly sampled half of the data points from each simulation between the first and last data point and fit the resulting dataset (including the first and last datapoint) a number of times -- and measured from this dataset a mean $n_{\infty}$ and a standard deviation $\sigma_{n_{\infty}}$ about that mean. We found that $\sigma_{n_{\infty}}$ was comparable to the $\Delta n_{\infty}$ given here and presented in the tables.}
\begin{equation}
\Delta n_{\infty} = \sqrt{2\left(\frac{\partial^2\chi^2}{\partial n_{\infty}^2}\right)^{-1}\chi^2}.
\end{equation}
This quantity represents the degree to which we have to change $n_{\infty}$ -- with all other parameters held fixed -- to correspondingly change the value of the $\chi^2$ by a factor of 2, and thus is a measure of the sensitivity of the goodness of our fit to the value of $n_{\infty}$. The last column of Table \ref{tab:1} gives the value of $\Delta n_{\infty}$ for the $1M_{\odot}$, ZAMS simulation with $\beta = 1.34$. Because $\Delta n_{\infty} \simeq 0.035$ for this specific simulation, we can only change $n_{\infty}$ by this amount before we significantly worsen the fit of our function to the data. The value of $n_{\infty}$ that we measure by this method is therefore a relatively robust quantity.

\begin{figure*}
	\includegraphics[width=0.505\textwidth]{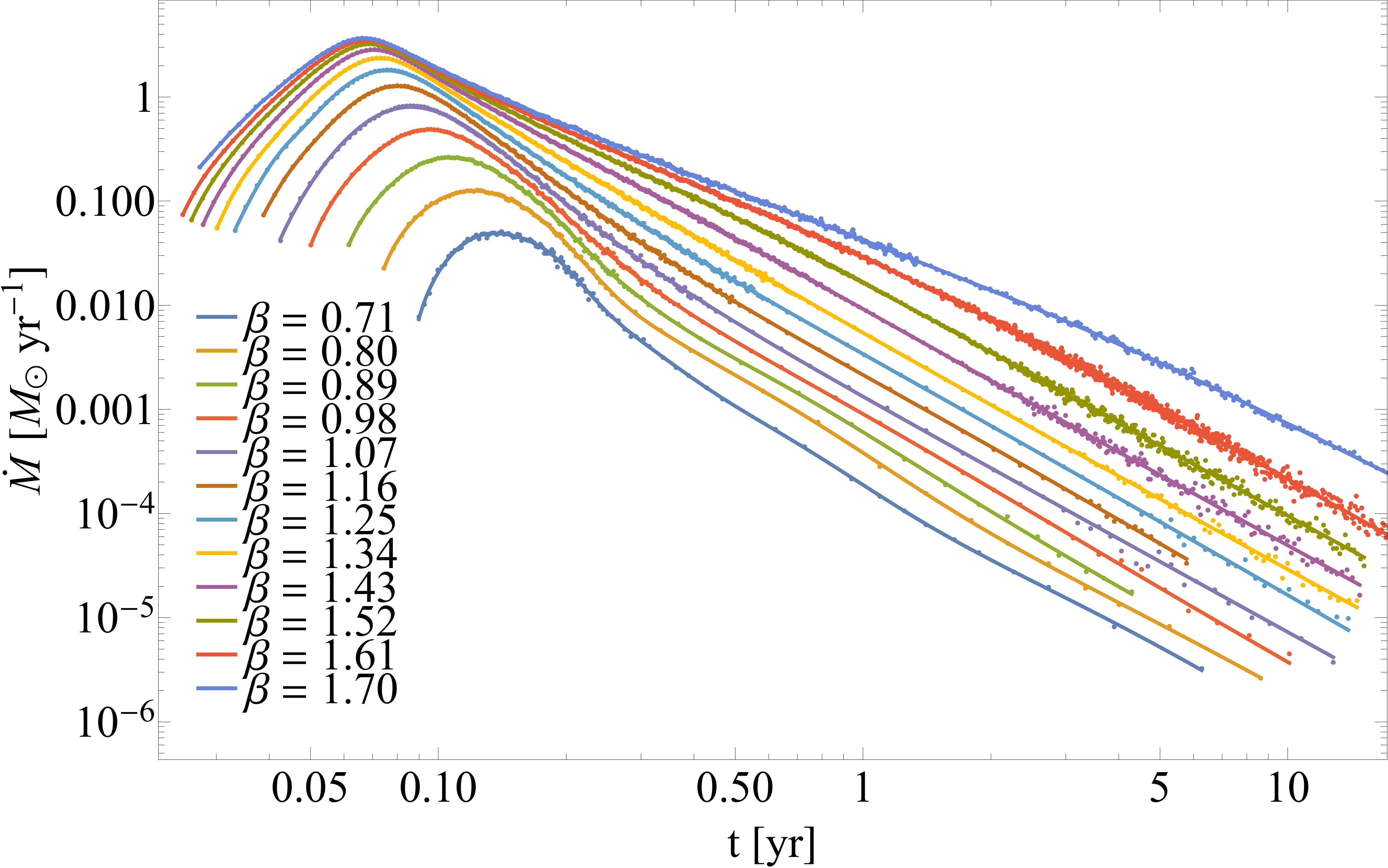}
	\includegraphics[width=0.485\textwidth]{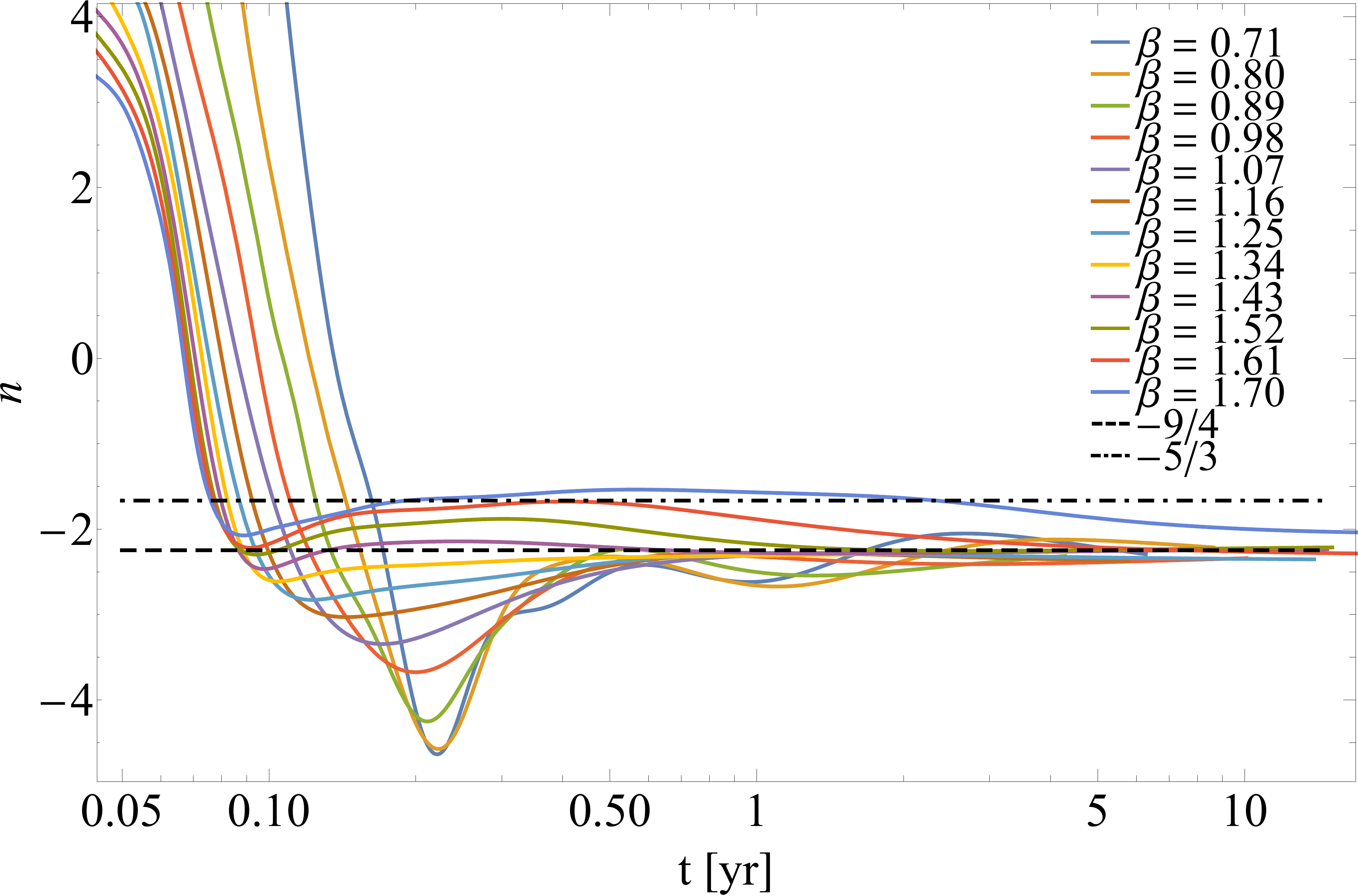}
	\includegraphics[width=0.505\textwidth]{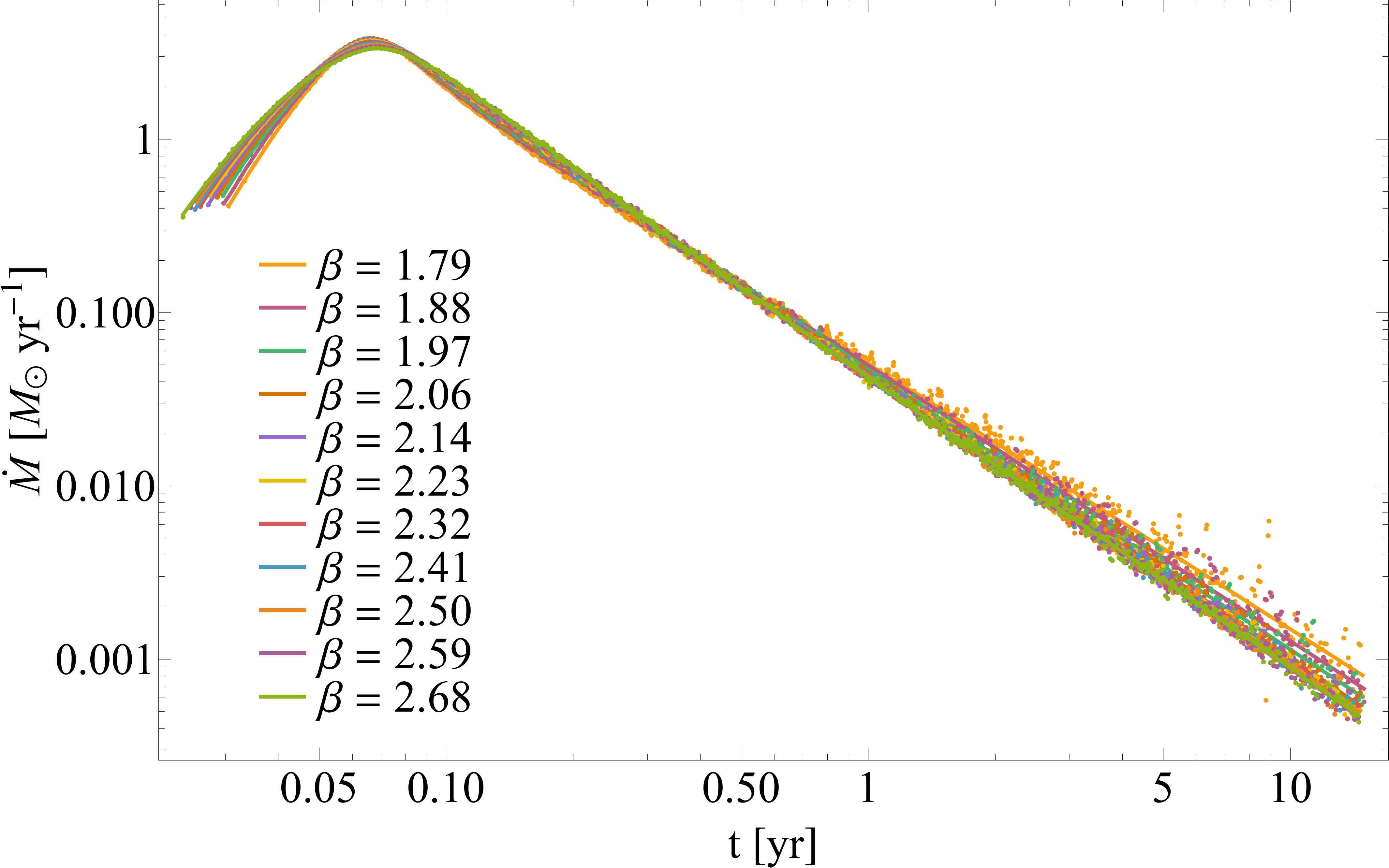}
	\includegraphics[width=0.485\textwidth]{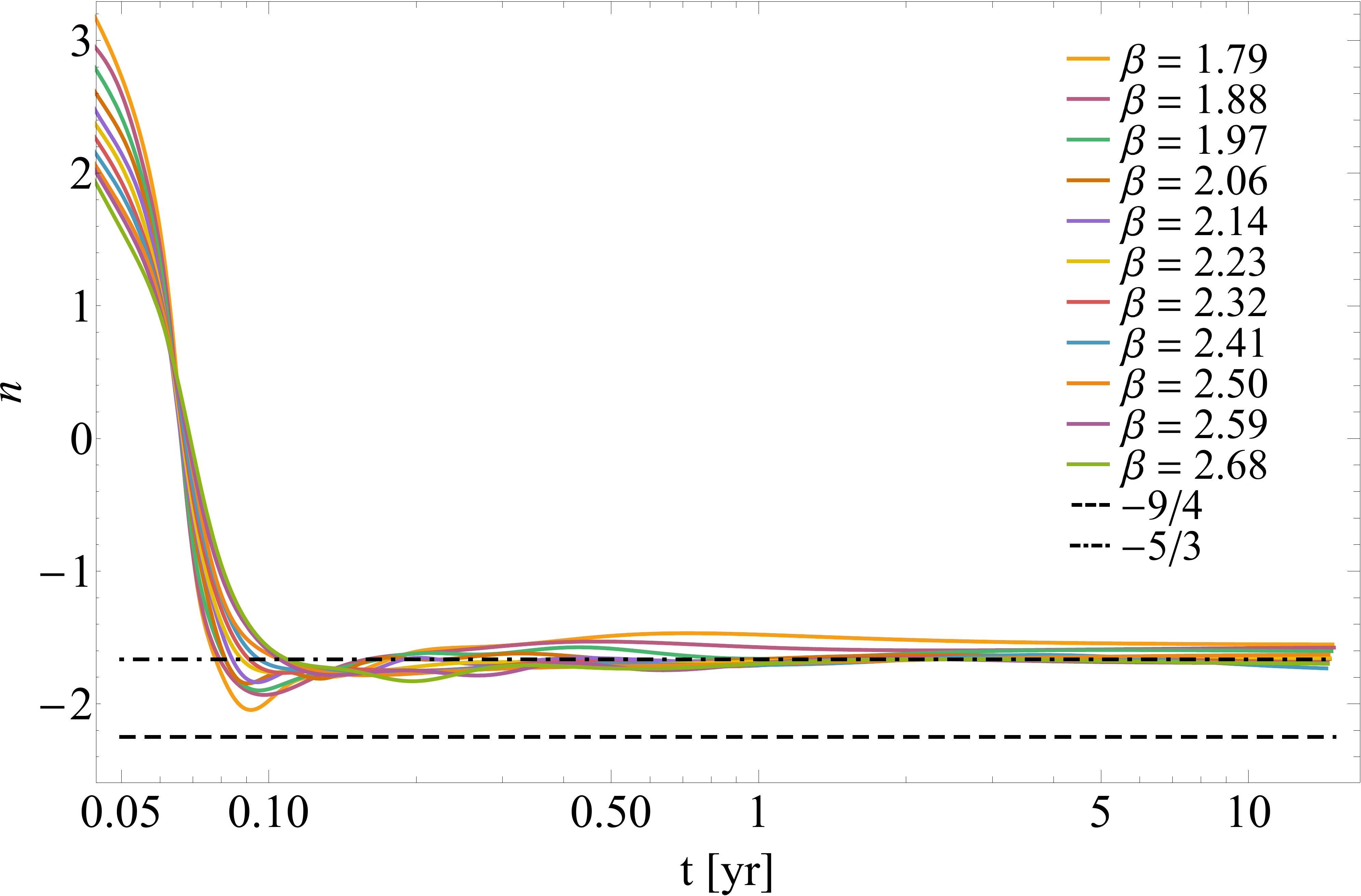}
	\caption{The fallback rate from the 1$M_{\odot}$ ZAMS disruptions with fits for partial disruptions and those with a zombie core (top-left) and full disruptions (bottom-left) and the instantaneous power-law index for partial/zombie-core disruptions (top-right) and full disruptions (bottom-right). The points in the left-hand panel give the results from the simulation, while the curves are from fits with Equation \eqref{Mdotfit}.}
	\label{fig8}
\end{figure*}

\begin{figure*}
	\includegraphics[width=0.505\textwidth]{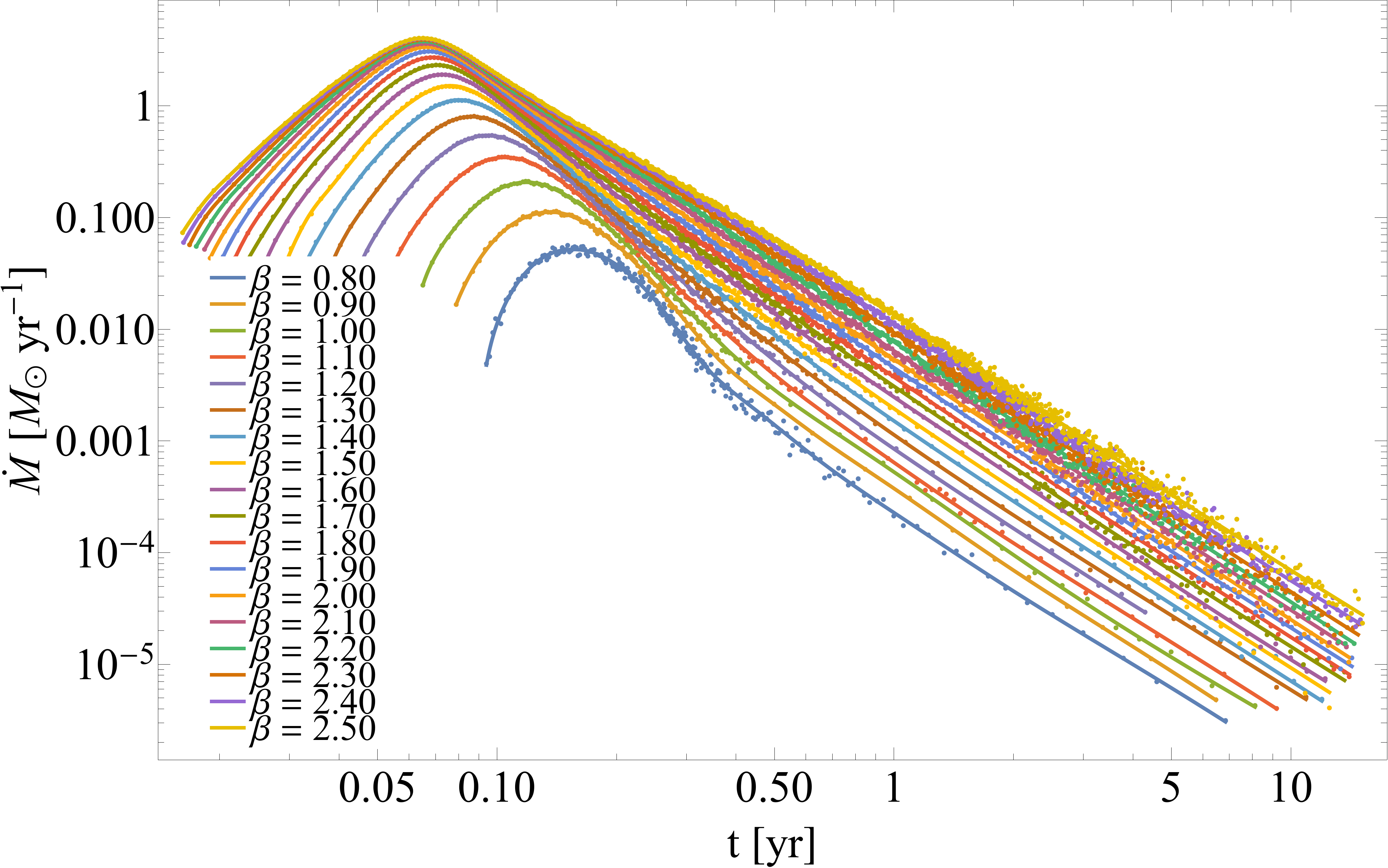}
	\includegraphics[width=0.485\textwidth]{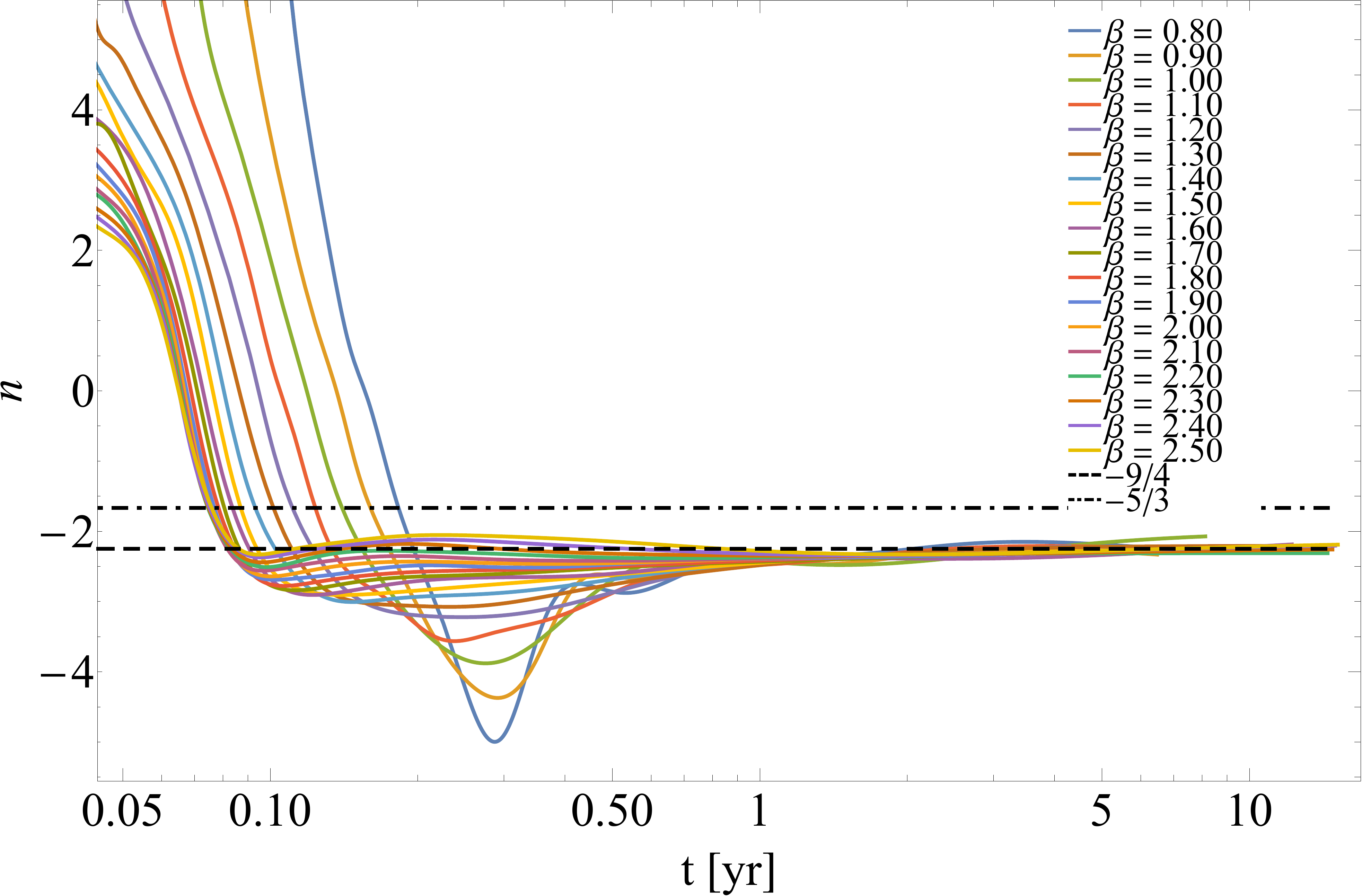}
	\includegraphics[width=0.505\textwidth]{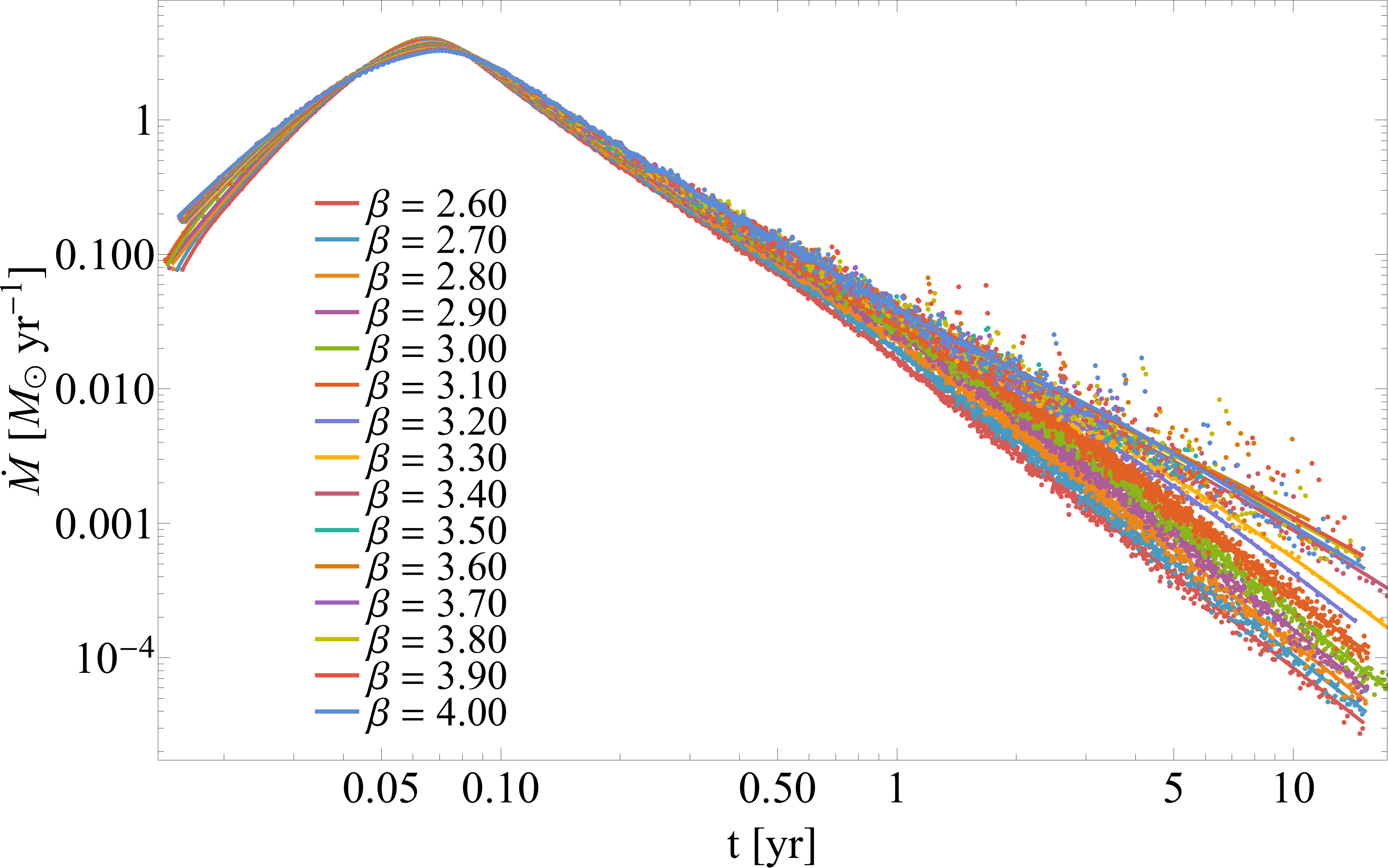}
	\includegraphics[width=0.485\textwidth]{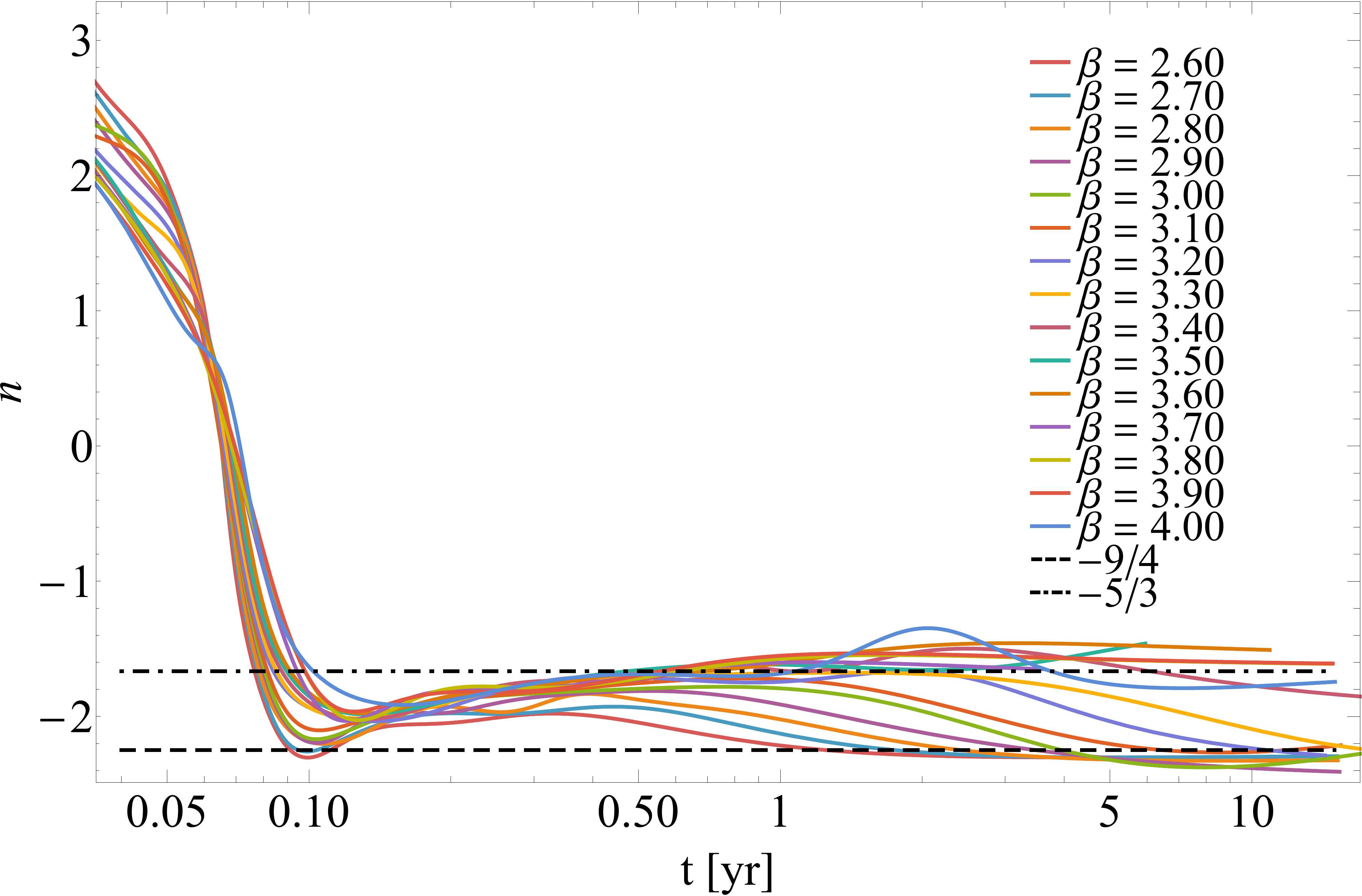}
	\caption{Same as for Figure \ref{fig8} but with the $1M_{\odot}$, MAMS star, except that we only include up to $\beta = 2.5$ in the top-left and top-right panels even though partial/zombie-core disruptions continue up until $\beta = 3.6$; we do this to maintain slightly less clutter in the figures.}
	\label{fig9}
\end{figure*}

\begin{figure*}
	\includegraphics[width=0.505\textwidth]{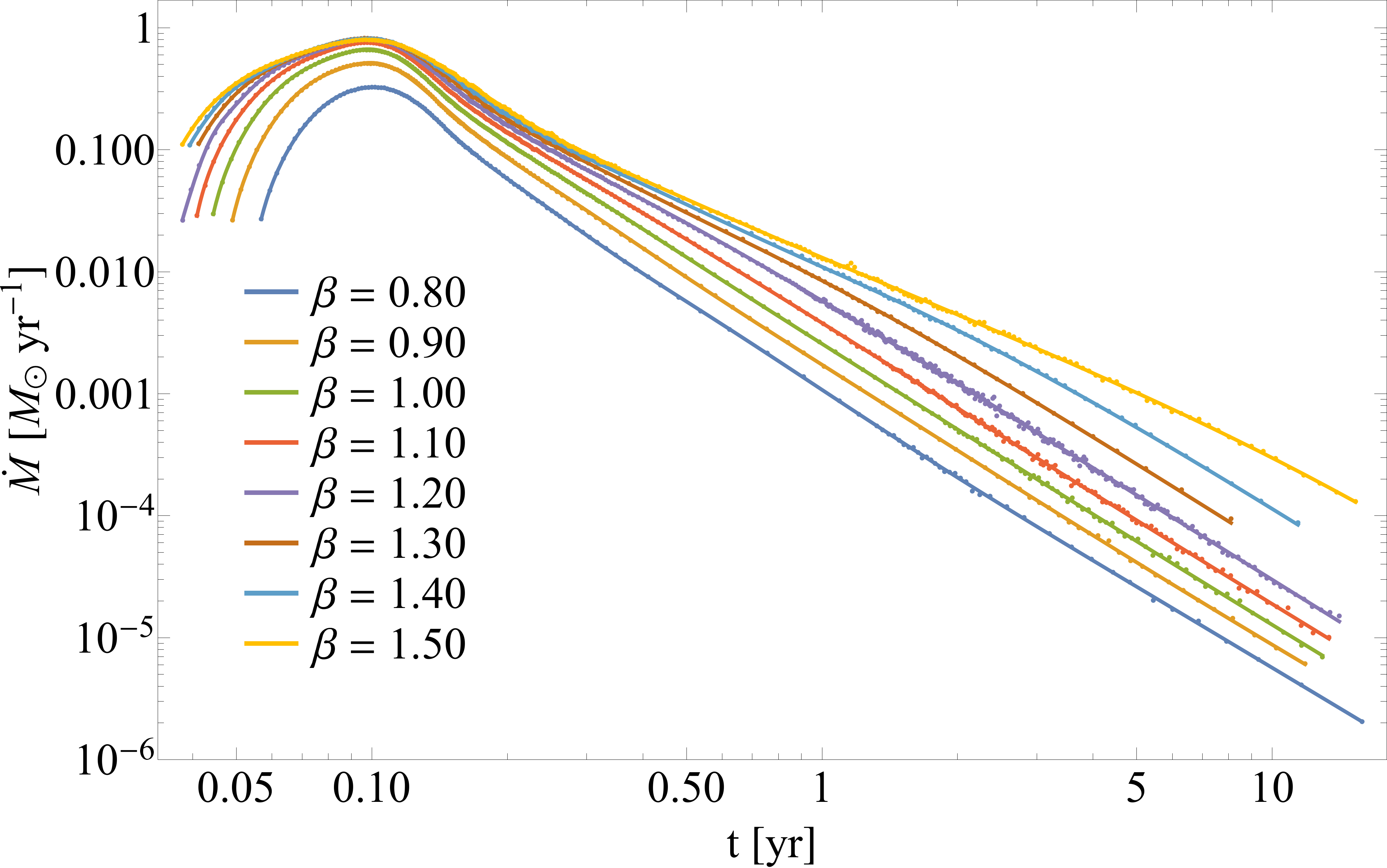}
	\includegraphics[width=0.485\textwidth]{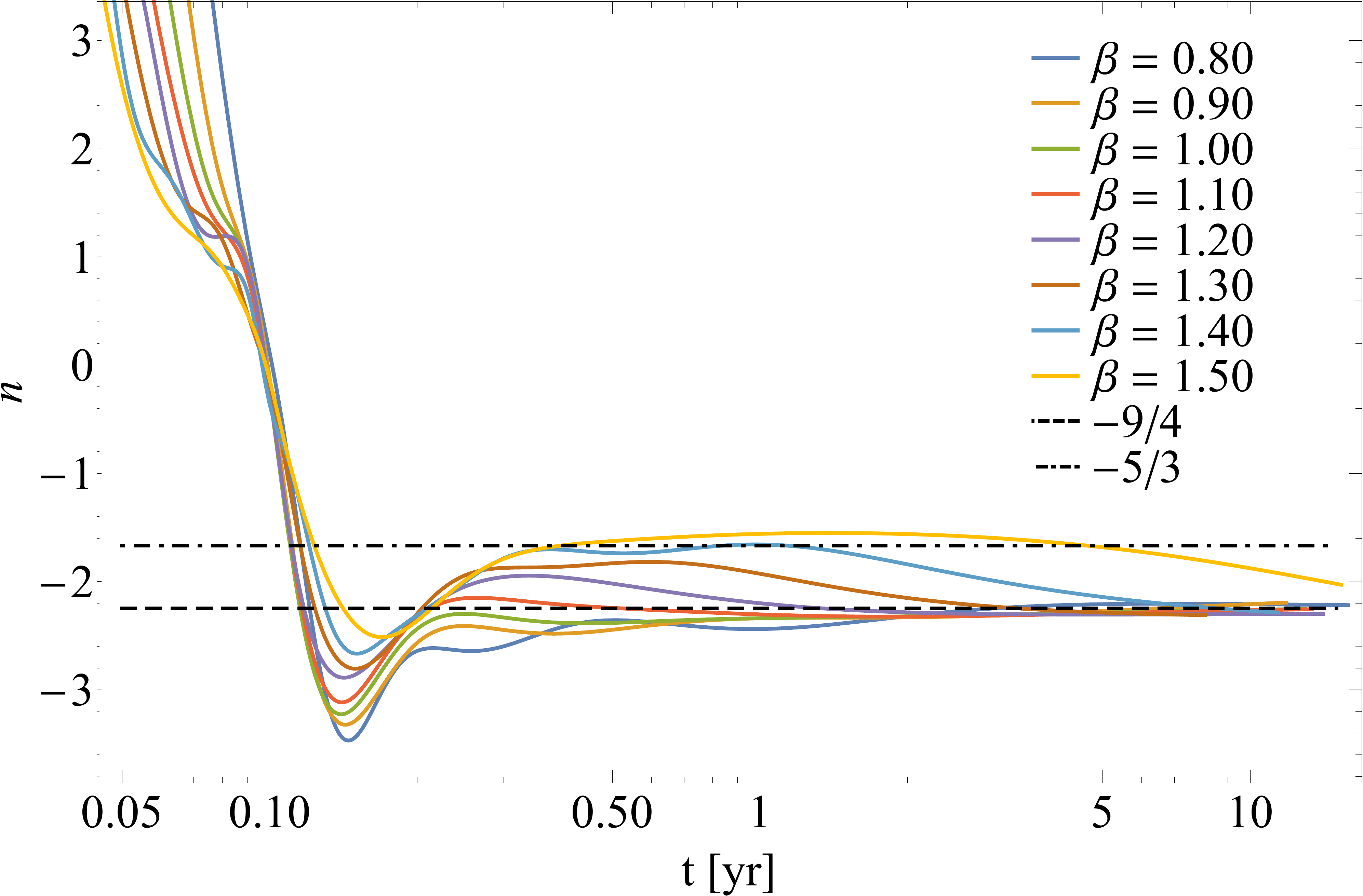}
	\includegraphics[width=0.505\textwidth]{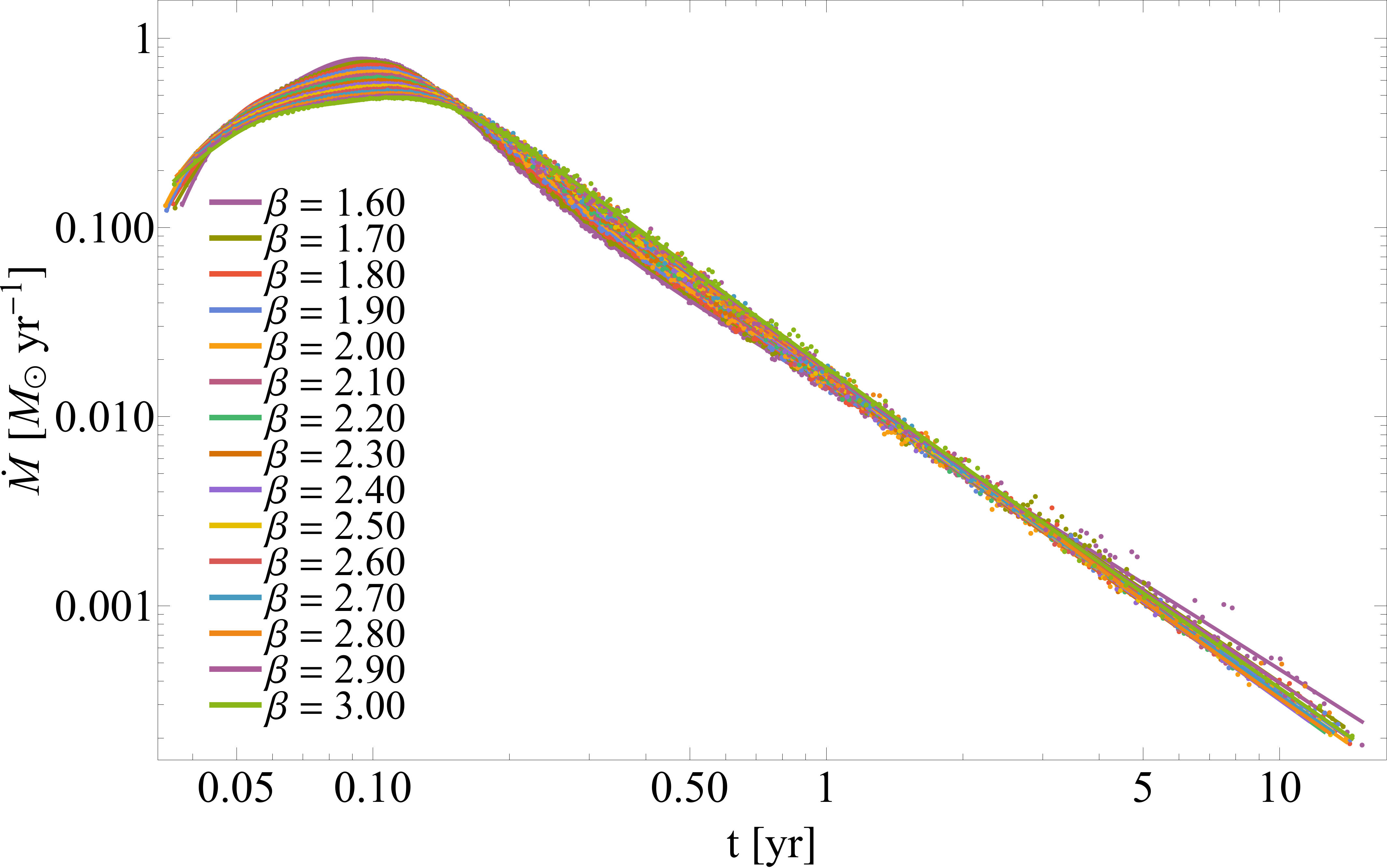}
	\includegraphics[width=0.485\textwidth]{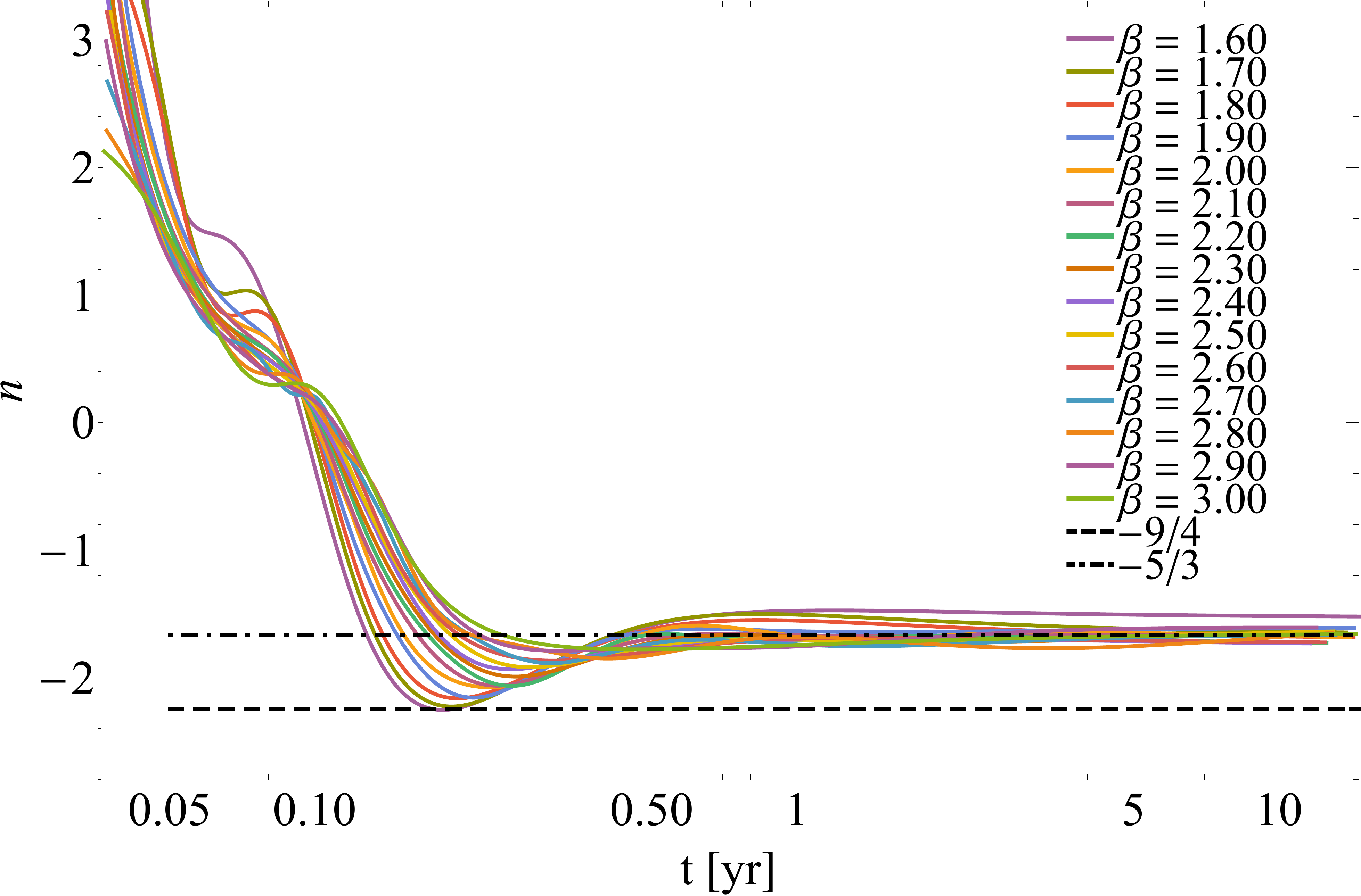}
	\caption{Same as for Figure \ref{fig8} but with the $0.3M_{\odot}$, MAMS star.}
	\label{fig10}
\end{figure*}

\begin{figure*}
	\includegraphics[width=0.505\textwidth]{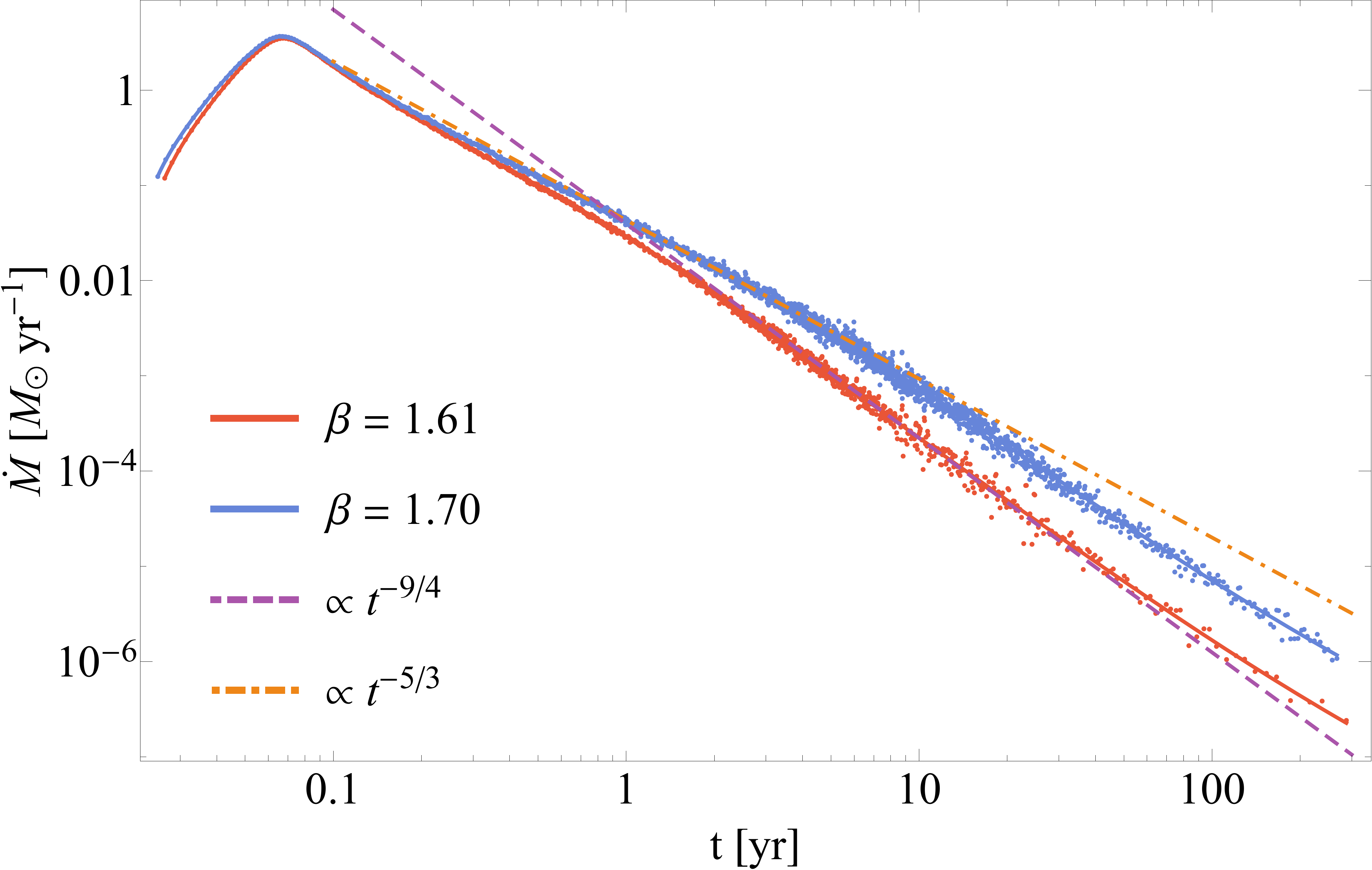}
	\includegraphics[width=0.485\textwidth]{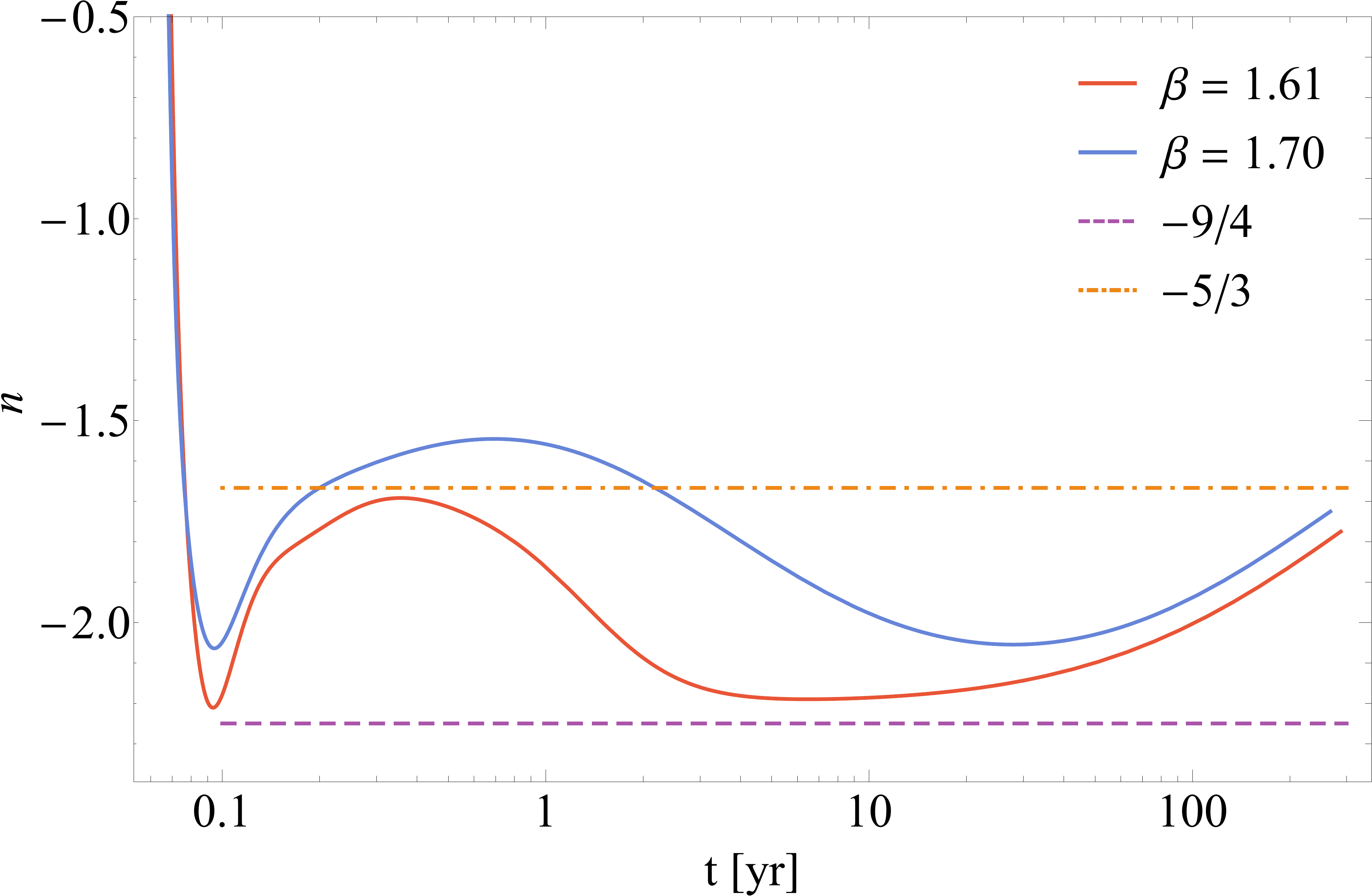}
	\caption{The fallback rate from the $\beta = 1.61$ (red) and $\beta = 1.70$ (blue) zombie disruptions of the $1M_{\odot}$ ZAMS star, where here we have run the simulations out to $\gtrsim 100$ years post-disruption. The left panel gives the fallback rate, while the right panel shows the instantaneous power-law index.}
	\label{fig11}
\end{figure*}

The parameters resulting from the fits to all of the fallback curves (i.e., for all $\beta$) are presented for the $1M_{\odot}$ ZAMS, $1M_{\odot}$ MAMS, and $0.3M_{\odot}$ MAMS in Tables \ref{tab:2}, \ref{tab:3}, and \ref{tab:4}, respectively. We note that the $\chi^2$ that results from fitting the fallback curves with $N_{\rm max} = 1$ (zero terms in the Pad\'e approximant), 3, or 5 are mostly of the order unity or below. Ordinarily this finding would suggest that our model is over-fitting the data. However, we emphasize that the number of data points for each simulation (of the order hundreds to thousands) is much greater than the number of parameters (4, 6, or 8) we are fitting to the data. Therefore, the small value of the $\chi^2$ implies that the data is very well-approximated by a function that rises as a power-law, decays as a power-law, and peaks in between -- as our function given in Equation \eqref{Mdotfit} does by construction. We also note that there is, for the majority of the simulations, a large reduction in the $\chi^2$ in going from 0 terms in the Pad\'e expansion ($N_{\rm max} = 1$) to two terms ($N_{\rm max} = 3$). This reduction in the $\chi^2$ occurs because by including two additional terms in the expansion we can exactly reconstruct the early and late-time rise and decay, \emph{and} the peak magnitude and the time of the peak. Further improvements in the fits are possible, and do occur, as $N_{\rm max}$ increases for $\beta \simeq \beta_{\rm c}$, as in these cases the evolution of the fallback curve is more complex (see Figure \ref{fig11} below and the corresponding discussion). Finally, we emphasize that \emph{even with no terms in the Pad\'e approximant} ($N_{\rm max} = 1$) the fit to the data is extremely good ($\chi^2 \ll N_{\rm points}$, where $N_{\rm points}$ is the number of data points in the simulation).

From the fits to the fallback curves we calculate the instantaneous power-law index of the fallback rate $n(t)$ for every simulation. The results for the $1M_{\odot}$, ZAMS star are shown in Figure \ref{fig8}, $1M_{\odot}$, MAMS in Figure \ref{fig9}, and $0.3M_{\odot}$ MAMS in Figure \ref{fig10}. In Figures \ref{fig8} and \ref{fig10} the top-left plot compares the data from the fallback rates (points) to the fits (curves) for all of the simulations that left a bound core, while the top-right panel gives the instantaneous power-law index for these same simulations. The bottom-left and bottom-right panels are the same as the top-left and top-right but for the simulations that resulted in the complete disruption of the star. For Figure \ref{fig9} we plot only simulations up to $\beta = 2.5$ in the top left and right panels even though partial disruptions continue up until $\beta = 3.6$; we do this to slightly ease the interpretation and appearance of this figure. 

From the combination of Figures \ref{fig8}, \ref{fig9}, and \ref{fig10}, it is apparent that the late-time behavior of the power-law index of \emph{most} of the fallback rates with surviving/zombie cores is to asymptote to a value that is close to $-9/4$, while the power-law indices of the simulations without cores is to asymptote to a value that is near $-5/3$. The first exception to this rule are the disruptions that form a zombie core, and are just below the critical $\beta$ for a full disruption, which are $\beta_{\rm c} \simeq 1.79$ for $1M_{\odot}$ ZAMS, $\beta_{\rm c} = 3.7$ for $1M_{\odot}$ MAMS, and $\beta_{\rm c} = 1.6$ for $0.3M_{\odot}$ MAMS. It is apparent from Figures \ref{fig8}, \ref{fig9}, and \ref{fig10} that these $n(t)$ curves initially follow $n(t) \simeq -5/3$ for an appreciable amount of time ($\sim$ years), then fall back below $-5/3$, but never quite steepen to $-9/4$. This behavior is reminiscent of that seen for the $\beta = 0.9$ disruption of a $1M_{\odot}$ star modeled as a $\gamma = 5/3$ polytrope investigated by \citet{Miles:2020aa}; in their case, the energy of the surviving core was slightly positive, which lessened the influence of the core on the fallback rate. Therefore, at very late times, the fallback rate of this simulation returned to $n(t) \simeq -5/3$ (see their Figure 2).

We find that the behavior of the cores very near the critical disruption threshold is similar to that found in \citet{Miles:2020aa}: the energy of the surviving core is slightly positive and its influence on the fallback rate lessens over time, resulting in a fallback rate that deviates from the expected $\propto t^{-9/4}$ scaling at sufficiently late times. This is shown explicitly in Figure \ref{fig11}, the left-hand panel of which gives the fallback rate from the $\beta = 1.61$ (red points) and $1.70$ (blue points) disruptions of the 1$M_{\odot}$ star, respectively, alongside their fits from the fitting function given in Equation \eqref{Mdotfit} (curves). The right panel shows the instantaneous power-law index obtained from the fits to the fallback data, which illustrates that the power-law index never quite steepens to $-9/4$ for each of these simulations. This figure also shows that, at timescales of $\sim$ hundreds of years, the power-law index eventually starts to return back to $-5/3$.

The energy of the surviving core that we measure from the $\beta = 1.61$ and $1.70$ simulations are, respectively, $\epsilon_{\rm c,\,1.61}/\Delta \epsilon = 3.68\times10^{-3}$ and $\epsilon_{\rm c,\, 1.70}/\Delta \epsilon = 6.21\times 10^{-3}$, where $\Delta \epsilon = GM_{\bullet}R_{\star}/R_{\rm t}^2$ is the energy of the most-unbound debris under the frozen-in approximation \citep{Lacy:1982aa}. We can estimate the time at which we expect the positive energy of the core to influence the fallback rate by noting that the equation of motion of the core is
\begin{equation}
\frac{1}{2}\left(\frac{dR_{\rm c}}{dt}\right)^{2}+\frac{1}{2}\frac{\ell_{\rm c}^2}{R_{\rm c}^2}-\frac{GM}{R_{\rm c}} = \epsilon_{\rm c},
\end{equation}
where $\ell_{\rm c}$ and $\epsilon_{\rm c}$ are the angular momentum and energy of the core, respectively. To leading order in $\ell_{\rm c}$ and $\epsilon_{\rm c}$ the solution to this equation is
\begin{equation}
R_{\rm c} \simeq R_0(t)\left(1+\frac{1}{5}\frac{\epsilon_{\rm c}}{GM}R_0-\frac{\ell_{\rm c}^2}{2GM}\right),
\end{equation}
where $R_0(t)$ is the zero-energy, zero-angular-momentum orbit that satisfies
\begin{equation}
R_0 = \left(\frac{3}{2}\sqrt{2GM}t\right)^{2/3}.
\end{equation}
The core will noticeably affect the fallback rate once $\epsilon_{\rm c}/(5GM)R_0(t) \simeq 1$, which, upon rearranging, shows that this occurs at a time of
\begin{equation}
T_{\epsilon} \simeq \frac{GM}{\epsilon_{\rm c}^{3/2}}.
\end{equation}
Inserting the energy of the surviving core from each simulation into this expression gives $T_{\epsilon, \, 1.61} \simeq 190$ years and $T_{\epsilon, \,1.70} \simeq 87$ years. This timescale is in rough agreement with the time at which we start to see the power-law index of the fallback rate from each of these simulations return to $\sim -5/3$, as evidenced by Figure \ref{fig11}.

We also note from Figures \ref{fig8} -- \ref{fig10} that the power-law index for disruptions that are at $\beta_{\rm c}$ and just above, and thus just barely disrupt the entire star, have power-law indices that stay above $-5/3$ at late times by a small but noticeable amount. This increase above the expected value arises from the fact that self-gravity in these simulations is trying to cause the disrupted debris stream to recollapse into a core, which results in a gradual increase (relative to the self-gravity-less case) in the amount of material near the marginally bound radius within the stream (from which material rains back onto the black hole at asymptotically late times). 

\section{Conclusions}
\label{conclusions}
In this paper we have presented numerical simulations of the tidal disruption of stars by a supermassive black hole. In particular we have used stars with accurate density profiles, corresponding to the $1M_\odot$ ZAMS, $1M_\odot$ MAMS and $0.3M_\odot$ MAMS stars modelled by \cite{Golightly:2019ab}, and simulated TDEs with the stars on parabolic orbits with pericentres that encompass partial and full disruptions. With these simulations, and a new methodology for providing analytical fits to the fallback rate data, we have established the following results:
\begin{itemize}
\item As predicted by \cite{Coughlin:2019aa}, there is a dichotomy in the late-time power-law indices for the fallback rates in TDEs, with full disruptions yielding $n_\infty \simeq -5/3$ and partial disruptions yielding $n_\infty \simeq -9/4$.
\item For orbits with $\beta \simeq \beta_{\rm c}$ it is possible for the central regions of the fully disrupted debris stream to re-collapse into a single, dominant, zombie core \citep[also seen in the simulations of ][]{Guillochon:2013aa,Golightly:2019ab}. For the cases presented here, the resulting core has a small positive orbital energy with respect to the black hole, resulting in the evolution of $n(t)$ from $\approx -9/4$ to $\approx -5/3$ on a timescale of $\sim$ 100 years. This highlights the need for long-duration simulations to capture the full behaviour of TDE debris streams (see Fig.~\ref{fig11}).
\item For some cases with $\beta \simeq \beta_{\rm c}$ it is possible that the central regions of the stream fragment into several zombie cores, rather than a single one (see Fig.~\ref{fig5}).
\item There is a complex relationship between the dependency of features in the fallback rates (e.g. rise time, time of peak and peak rate) on the impact parameter $\beta$ and the stellar properties (see Fig.~\ref{fig3}). This is evidenced by the variation in time of peak with $\beta$ for the $1M_\odot$ stars, and the relative constancy of the time of peak for the $0.3M_\odot$ star.
\item We have shown that for impact parameters up to at least twice the critical value at which the star is fully disrupted, the effects of self-gravity of the debris stream can manifest in the fallback rate. This was particularly evident in the $0.3M_\odot$ MAMS simulation with $\beta=3$, for which $\beta_{\rm c} \approx 1.5$ (shown in Fig.~\ref{fig4}).
\item We have also found that it is possible for the surviving stellar core in a partial TDE to acquire a circumstellar disc of material (see Fig.~\ref{fig6}). This disc is primarily formed from material that is ejected from the equator of the core, which is initially rotating above the critical rotation rate for equilibrium. The disc is also fed over time from the portion of the debris stream that is bound to the core. These cores, which exist on unbound orbits with respect to the central black hole, may make their way into the galaxy as rapidly rotating objects with potentially observable circumstellar disc emission.
\item Finally, we have presented a simple analytical function that can be used to fit TDE fallback rates with good fits achieved with relatively few parameters. We expect that this function (equation~\ref{Mdotfit}) may have wide application in transient astrophysics where variations in a measured quantity vary from one power-law to another in time.
\end{itemize}

\begin{acknowledgments}
C.J.N acknowledges funding from the European Union’s Horizon 2020 research and innovation program under the Marie Sk\l{}odowska-Curie grant agreement No 823823 (Dustbusters RISE project). E.R.C.~acknowledges support from the National Science Foundation through grant AST-2006684. P.R.M acknowledges support from the Syracuse University HTC Campus Grid and NSF award ACI-1341006. Some of this research used the ALICE High Performance Computing Facility at the University of Leicester. Some of this work was performed using the DiRAC Data Intensive service at Leicester, operated by the University of Leicester IT Services, which forms part of the STFC DiRAC HPC Facility (\url{www.dirac.ac.uk}). The equipment was funded by BEIS capital funding via STFC capital grants ST/K000373/1 and ST/R002363/1 and STFC DiRAC Operations grant ST/R001014/1. DiRAC is part of the National e-Infrastructure. We used {\sc splash} \citep{Price:2007aa} for some of the figures.
\end{acknowledgments}

\bibliographystyle{aasjournal}
\bibliography{nixon}

\appendix
\section{Fitted parameters}
Here we present tabulated data from the simulations to un-clutter the main body of the text. Table \ref{tab:1} gives the fitting parameters from the fitting function defined in Equation \eqref{Mdotfit} for the $\beta = 1.34$ disruption of the 1$M_{\odot}$, ZAMS star for different values of $N_{\rm max}$ (where there are $N_{\rm max}-1$ coefficients in the Pad\'e approximant in Equation \ref{Mdotfit}). Tables \ref{tab:2}, \ref{tab:3}, and \ref{tab:4} give the best-fit parameters that minimize the $\chi^2$ for the $1M_{\odot}$ ZAMS, $1M_{\odot}$ MAMS, and $0.3M_{\odot}$ MAMS star, respectively, for all of the simulations. The details of what each parameter means is given in the caption of the table.

\begin{table*}[h!]
\begin{center}
\begin{tabular}{|c|c|c|c|c|c|c|c|c|c|c|c|c|c|c|c|}
\hline
$N_{\rm max}$ & $a$ & $m$ & $b$ & $n_{\infty}$ & $c_1$ & $c_2$ & $c_3$ & $c_4$ & $c_5$ & $c_6$ & $c_7$ & $c_8$ & $c_9$ & $\chi^2$ & $\Delta n_{\infty}$ \\ 
\hline
1 & 10.0 & 5.87 & 0.470 & -2.32 & N/A & N/A & N/A & N/A & N/A & N/A & N/A & N/A & N/A &  1.71 & 0.0450 \\
\hline
2 & 5.71 & 5.60 & 0.693 & -2.23 & 1.16 & N/A & N/A & N/A & N/A & N/A & N/A & N/A & N/A &  0.887 & 0.0324 \\
\hline
3 & 5.04 & 5.33 & 0.531 & -2.29 & 0.932 & -0.119 & N/A & N/A & N/A & N/A & N/A & N/A & N/A &  0.652 & 0.0279 \\
\hline
4 & 2.77 & 5.18 & 0.706 & -2.23 & 4.14 & -3.69 & 1.76 & N/A & N/A & N/A & N/A & N/A & N/A &  0.5985 & 0.0266 \\
\hline
5 & 2.40 & 5.17 & 0.601 & -2.26 & 5.94 & -6.30 & 2.13 & 0.304 & N/A & N/A & N/A & N/A & N/A &  0.5983 & 0.0266\\
\hline
6 & 0.963 & 5.49 & 0.764 & -2.22 & 38.9 & -69.3 & 48.2 & -15.8 & 3.23 & N/A & N/A & N/A & N/A &  0.584 & 0.0263 \\
\hline
7 & 2884 & 7.97 & 0.597 & -2.26 & 1.78 & -18.8 &32.9 & -21.0 & 5.41 & 0.0350 & N/A & N/A & N/A &  0.485 & 0.0240 \\
\hline
8 & 8453 & 11.7 & 0.644 & -2.25 & 0.820 & -8.60 & 18.6 & -12.5 & 4.42 & -1.52 & 1.00 & N/A & N/A &  0.472 & 0.0236 \\
\hline
9 & 431709 & 13.6 & 0.701 & -2.23 & -9.69 & 42.4 & -103 & 144 & -111 & 48.0 & -11.7 & -2.26 & N/A & 0.464 & 0.0235 \\
\hline
10 & 35133 & 9.93 & 0.634 & -2.25 & -13.4 & 72.4 & -202 & 314 & -277 & 138 & -36.4 & 3.83 & 0.707 & 0.462  & 0.0235 \\
\hline
\end{tabular}
\caption{For the $1M_{\odot}$, ZAMS star with $\beta = 1.35$, the coefficients in the fitted function as we increase the number of terms in the Pad\'e approximant (the ratio of polynomials in Equation \ref{Mdotfit}). The quantities $a$ and $m$ are the normalization and power-law index of the initial rise of the fallback rate, $b$ and $n_{\infty}$ are the normalization and power-law index of the decay of the fallback rate, and $c_1$, $c_2$, \ldots, $c_9$ are the coefficients in the polynomial expansion of the Pad\'e approximant. The $\chi^2$ of the fit is given by the penultimate column (here there are 346 data points from which the $\chi^2$ was constructed), and $\Delta n_{\infty}$ is the change in $n_{\infty}$ that would result in an increase of the $\chi^2$ by a factor of two. }
\label{tab:1}
\end{center}
\end{table*}

\begin{table*}
\begin{center}
\begin{tabular}{|c|c|c|c|c|c|c|c|c|c|c|c|}
\hline
$\beta$ & $n_{\infty,1}$ & $n_{\infty,3}$ & $n_{\infty,5}$ & $n_{10,3}$ & $n_{10,5}$ & $\chi_1^2 (N_{\rm points})$ & $\chi_3^2 $ & $\chi_5^2$ & $\Delta n_{\infty,1}$ & $\Delta n_{\infty,3}$ & $\Delta n_{\infty,5}$ \\ 
\hline
0.715  & -2.77 & -2.41 & -2.16 & -2.37 & -2.24 & 4.45 (100) & 0.810 & 0.506 & 0.313 & 0.133 & 0.105 \\ 
\hline
0.803  & -2.71 & -2.48 & -2.16  & -2.45 & -2.27 & 3.53 (82) & 0.479 & 0.221 & 0.256 & 0.0943 & 0.0645 \\ 
\hline
0.894  & -2.67 & -2.36 & -1.96  & -2.33 & -2.16  & 6.92 (113) & 0.707 & 0.327 & 0.267 & 0.0858 & 0.0588 \\ 
\hline
0.983  & -2.63 & -2.34 & -2.52  & -2.33 & -2.48  & 7.86 (148)& 0.477 & 0.243& 0.236 & 0.0581 & 0.0414 \\ 
\hline
1.07 & -2.55 & -2.26 & -2.26  & -2.26 & -2.27  & 8.51 (186)& 0.643 & 0.488 & 0.182 & 0.0502 & 0.0438 \\ 
\hline
1.16  & -2.53 & -2.29 & -2.41  & -2.29 & -2.39  & 5.27 (215) & 0.197 & 0.192 & 0.144 & 0.0280 & 0.0278 \\ 
\hline
1.25  & -2.41 & -2.28 & -2.27  & -2.28 & -2.27  & 4.15 (268) & 0.503 & 0.503 & 0.0939 & 0.0327 & 0.0327 \\ 
\hline
1.34  & -2.33 & -2.29 & -2.26  & -2.29 & -2.27  & 1.71 (346) & 0.652 & 0.598 & 0.0450 & 0.0279 & 0.0266 \\ 
\hline
1.43  & -2.25 & -2.30 & -2.27  & -2.30 & -2.27  & 2.52 (514) & 2.15 & 2.08 & 0.0366 & 0.0338 & 0.0332 \\ 
\hline
1.52 & -2.14 & -2.28 & -2.31  & -2.27 & -2.29  & 8.52 (741) & 2.76 & 2.75 & 0.0498 & 0.0282 & 0.0282 \\ 
\hline
1.61 & -1.97 & -2.16 & -2.28  & -2.15 & -2.24 & 27.7 (1230) & 8.72 & 5.89 & 0.0621 & 0.0348 & 0.0286 \\ 
\hline
1.70 & -1.67 & -1.77 & -1.89  & -1.76 & -1.85  & 4.48 (548) & 1.74 & 1.04 & 0.0471 & 0.0294 & 0.0226 \\ 
\hline
1.79 & -1.53 & -1.55 & -1.58  & -1.54 & -1.56  & 18.2 (1030)& 15.9 & 15.7 & 0.0641 & 0.0598 & 0.0595 \\ 
\hline
1.88 & -1.58 & -1.61 & -1.60  & -1.60 & -1.59  & 5.49 (502)& 4.48 & 4.39 & 0.0482 & 0.0436 & 0.0431 \\ 
\hline
1.97 & -1.63 & -1.64 & -1.59  & -1.63 & -1.60  & 3.93 (484) & 3.39 & 3.22 & 0.0429 & 0.0399 & 0.0387 \\ 
\hline
2.06 & -1.66 & -1.67 & -1.64  & -1.67 & -1.64  & 2.06 (476)& 1.66 & 1.54 & 0.0318 & 0.0286 & 0.0276 \\ 
\hline
2.14 & -1.68 & -1.68 & -1.66  & -1.67 & -1.66  & 1.72 (486)& 1.34 & 1.24 & 0.0286 & 0.0252 & 0.0243 \\ 
\hline
2.23 & -1.68 & -1.67 & -1.65  & -1.67 & -1.65  & 1.76 (508)& 1.31 & 1.21 & 0.0281 & 0.0243 & 0.0233 \\ 
\hline
2.32 & -1.69 & -1.67 & -1.65  & -1.67 & -1.66  & 1.70 (496) & 1.22 & 1.15 & 0.0280 & 0.0238 & 0.0231 \\ 
\hline
2.41 & -1.70 & -1.67 & -1.65  & -1.67 & -1.66 & 1.76 (504) & 1.24 & 1.18 & 0.0284 & 0.0239 & 0.0232 \\ 
\hline
2.50 & -1.70 & -1.66 & -1.64  & -1.66 & -1.65  & 2.29 (530) & 1.64 & 1.58 & 0.0317 & 0.0269 & 0.0263 \\ 
\hline
2.59 & -1.71 & -1.68 & -1.69  & -1.68 & -1.68  & 1.79 (496) & 1.31 & 1.28 & 0.0290 & 0.0247 & 0.0246 \\ 
\hline
2.68 & -1.71 & -1.67 & -1.66  & -1.68 & -1.66 & 2.26 (553) & 1.70 & 1.67 & 0.0307 & 0.0266 & 0.0264 \\ 
\hline
\end{tabular}
\caption{The parameters for each simulation for the $1M_{\odot}$, ZAMS star: the $\beta$ of the encounter, $n_{\infty, 1}$, $n_{\infty, 3}$, and $n_{\infty, 5}$ the asymptotic power-law coefficient with $N_{\rm max} = 1$, 3, and 5 respectively, $n_{10,3}$ and $n_{10,5}$ the instantaneous power-law index of the fallback at ten years with $N_{\rm max} = 3$ and 5 respectively, $\chi^2_1$, $\chi^2_3$, and $\chi^2_5$ the $\chi^2$ with $N_{\rm max} = 1$, 3, and 5 respectively ($N_{\rm points}$ is the number of data points in each fallback curve used to construct the $\chi^2$), and $\Delta n_{\infty,1}$, $\Delta n_{\infty,3}$, and $\Delta n_{\infty,5}$ the change in $n_{\infty}$ required to change the $\chi^2$ by a factor of 2 with $N_{\rm max} = 1$, 3, and 5, respectively. }
\label{tab:2}
\end{center}
\end{table*}

\begin{table*}
\begin{center}
\begin{tabular}{|c|c|c|c|c|c|c|c|c|c|c|c|c|}
\hline
$\beta$ & $n_{\infty,1}$ & $n_{\infty,3}$ & $n_{\infty,5}$ & $n_{10,3}$ & $n_{10,5}$ & $\chi_1^2 (N_{\rm points})$ & $\chi_3^2 $ & $\chi_5^2$ & $\Delta n_{\infty,1}$ & $\Delta n_{\infty,3}$ & $\Delta n_{\infty,5}$ \\ 
\hline
0.8  & -2.94 & -2.50 & -2.29 & -2.37 & -2.36 & 19.9 (235) & 5.38 & 4.24 & 0.414 & 0.216 & 0.195 \\ 
\hline
0.9 & -2.84 & -2.44 & -2.20  & -2.45 & -2.30 & 5.04 (103) & 0.517 & 0.209 & 0.325 & 0.104 & 0.0686 \\ 
\hline
1.0  & -2.78 & -2.29 & -1.94  & -2.33 & -2.16  & 6.45 (131) & 0.528 & 0.259 & 0.270 & 0.0772 & 0.0559 \\ 
\hline
1.1  & -2.78 & -2.17 & -2.26  & -2.33 & -2.25  & 7.43 (166)& 0.407 & 0.300& 0.221 & 0.0515 & 0.0443 \\ 
\hline
1.2 & -2.82 & -2.02 & -1.99  & -2.26 & -2.06  & 4.32 (180)& 0.376 & 0.256 & 0.187 & 0.0552 & 0.0455 \\ 
\hline
1.3  & -2.66 & -2.16 & -2.07  & -2.29 & -2.14  & 6.81 (209) & 0.449 & 0.439 & 0.173 & 0.0443 & 0.0438 \\ 
\hline
1.4  & -2.60 & -2.24 & -2.19  & -2.28 & -2.22  & 6.43 (236) & 0.477 & 0.465 & 0.141 & 0.0383 & 0.0378 \\ 
\hline
1.5  & -2.58 & -2.35 & -2.42  & -2.29 & -2.41  & 5.09 (264) & 0.852 & 0.802 & 0.111 & 0.0457 & 0.0444 \\ 
\hline
1.6  & -2.53 & -2.32 & -2.30  & -2.30 & -2.31  & 4.50 (296) & 0.766 & 0.654 & 0.0919 & 0.0379 & 0.0351 \\ 
\hline
1.7 & -2.48 & -2.34 & -2.22  & -2.27 & -2.25 & 5.46 (366) & 1.72 & 1.49 & 0.0755 & 0.0423 & 0.0395 \\ 
\hline
1.8 & -2.46 & -2.34 & -2.28  & -2.15 & -2.29 & 3.80 (365) & 0.953 & 0.739 & 0.0668 & 0.0335 & 0.0296 \\ 
\hline
1.9 & -2.43 & -2.35 & -2.29  & -1.76 & -2.30  & 3.54 (421) & 1.50 & 1.24 & 0.0550 & 0.0358 & 0.0325 \\ 
\hline
2.0 & -2.40 & -2.35 & -2.30  & -1.54 & -2.31  & 3.13 (472)& 1.95 & 1.66 & 0.0460 & 0.0362 & 0.0335 \\ 
\hline
2.1 & -2.38 & -2.35 & -2.28  & -1.60 & -2.29  & 4.70 (592)& 4.11 & 3.75 & 0.0403 & 0.0437 & 0.0417 \\ 
\hline
2.2 & -2.35 & -2.34 & -2.30  & -1.63 & -2.31  & 4.09 (652) & 4.00 & 3.69 & 0.0434 & 0.0399 & 0.0383 \\ 
\hline
2.3 & -2.32 & -2.33 & -2.30  & -1.67 & -2.31  & 5.87 (758)& 5.70 & 5.40 & 0.0450 & 0.0429 & 0.0417 \\ 
\hline
2.4 & -2.28 & -2.32 & -2.30  & -1.67 & -2.30  & 7.15 (811)& 6.31 & 6.09 & 0.0475 & 0.0423 & 0.0416 \\ 
\hline
2.5 & -2.24 & -2.29 & -2.32  & -1.67 & -2.31  & 9.76 (932)& 7.30 & 7.25 & 0.0501 & 0.0412 & 0.0410 \\ 
\hline
2.6 & -2.20 & -2.26 & -2.35  & -1.67 & -2.34  & 13.0 (1055) & 7.49 & 7.49 & 0.0564 & 0.0379 & 0.0379 \\ 
\hline
2.7 & -2.15 & -2.20 & -2.37  & -1.67 & -2.35 & 19.6 (1203) & 10.0 & 9.61 & 0.0578 & 0.0403 & 0.0395 \\ 
\hline
2.8 & -2.10 & -1.66 & -2.38  & -1.66 & -2.34  & 23.2 (1280) & 9.78 & 8.52 & 0.0629 & 0.0376 & 0.0351 \\ 
\hline
2.9 & -2.05 & -1.68 & -2.41  & -1.68 & -2.36  & 33.5 (1468) & 14.7 & 10.9 & 0.0684 & 0.0417 & 0.0361 \\ 
\hline
3.0 & -1.98 & -1.67 & -2.49  & -1.68 & -2.38 & 46.6 (1650) & 23.6 & 15.8 & 0.0307 & 0.0488 & 0.0406 \\ 
\hline
3.1 & -1.92 & -1.94 & -2.37  & -2.06 & -2.15 & 48.2 (1854) & 28.1 & 20.7 & 0.0641 & 0.0489 & 0.0426 \\ 
\hline
3.2 & -1.80 & -1.88 & -2.28  & -1.85 & -2.00 & 3.55 (432) & 1.58 & 1.01 & 0.0519 & 0.0346 & 0.0320 \\ 
\hline
3.3 & -1.76 & -1.81 & -1.95  & -1.79 & -1.87 & 4.41 (477) & 2.42 & 2.12 & 0.0535 & 0.0396 & 0.0380 \\ 
\hline
3.4 & -1.68 & -1.60 & -1.58  & -1.60 & -1.59 & 12.1 (641) & 7.39 & 7.29 & 0.0750 & 0.0587 & 0.0583 \\ 
\hline
3.5 & -1.70 & -1.64 & -1.62  & -1.64 & -1.62 & 8.69 (635) & 5.44 & 5.37 & 0.0704 & 0.0557 & 0.0553 \\ 
\hline
3.6 & -1.64 & -1.47 & -1.40  & -1.48 & -1.43 & 25.2 (663) & 18.2 & 17.9 & 0.112 & 0.0950 & 0.0943 \\ 
\hline
3.7 & -1.73 & -1.65 & -1.60  & -1.65 & -1.61 & 5.60 (576) & 3.66 & 3.59 & 0.0675 & 0.0546 & 0.0540 \\ 
\hline
3.8 & -1.65 & -1.56 & -1.55  & -1.56 & -1.55 & 22.2 (668) & 17.6 & 17.5 & 0.102 & 0.0905 & 0.0904 \\ 
\hline
3.9 & -1.64 & -1.51 & -1.47  & -1.52 & -1.49 & 25.2 (628) & 19.9 & 19.7 & 0.109 & 0.0966 & 0.0963 \\ 
\hline
4.0 & -1.66 & -1.60 & -2.00  & -1.60 & -1.85 & 20.5 (604) & 17.1 & 16.4 & 0.0986 & 0.0901 & 0.115 \\ 
\hline
\end{tabular}
\caption{The parameters for each simulation for the $1M_{\odot}$, MAMS star; parameters are the same as those in Table \ref{tab:2}.}
\label{tab:3}
\end{center}
\end{table*}

\begin{table*}
\begin{center}
\begin{tabular}{|c|c|c|c|c|c|c|c|c|c|c|c|c|}
\hline
$\beta$ & $n_{\infty,1}$ & $n_{\infty,3}$ & $n_{\infty,5}$ & $n_{10,3}$ & $n_{10,5}$ & $\chi_1^2 (N_{\rm points})$ & $\chi_3^2 $ & $\chi_5^2$ & $\Delta n_{\infty,1}$ & $\Delta n_{\infty,3}$ & $\Delta n_{\infty,5}$ \\ 
\hline
0.8  & -2.46 & -2.28 & -2.06 & -2.28 & -2.16 & 2.41 (142) & 0.472 & 0.166 & 0.102 & 0.0451 & 0.0267 \\ 
\hline
0.9 & -2.42 & -2.33 & -2.12  & -2.33 & -2.20 & 2.17 (156) & 0.795 & 0.197 & 0.0915 & 0.0553 & 0.0276 \\ 
\hline
1.0  & -2.37 & -2.34 & -2.16  & -2.34 & -2.23  & 2.12 (219) & 1.10 & 0.319 & 0.0658 & 0.0474 & 0.0256 \\ 
\hline
1.1  & -2.30 & -2.36 & -2.20  & -2.35 & -2.25  & 2.14 (299)& 1.49 & 0.390& 0.0505 & 0.0421 & 0.0216 \\ 
\hline
1.2 & -2.19 & -2.35 & -2.38  & -2.34 & -2.36  & 4.30 (456)& 2.10 & 0.767 & 0.0532 & 0.0372 & 0.0225 \\ 
\hline
1.3  & -2.02 & -2.24 & -2.38  & -2.22 & -2.32  & 2.46 (192) & 0.238 & 0.213 & 0.107 & 0.0332 & 0.0321 \\ 
\hline
1.4  & -1.86 & -2.03 & -2.08  & -2.00 & -2.04  & 3.46 (219) & 0.401 & 0.157 & 0.103 & 0.0349 & 0.0218 \\ 
\hline
1.5  & -1.71 & -1.72 & -1.85  & -1.70 & -1.78  & 4.94 (273) & 0.446 & 0.412 & 0.0914 & 0.0273 & 0.0272 \\ 
\hline
1.6  & -1.73 & -1.55 & -1.54  & -1.54 & -1.53  & 9.83 (392) & 1.54 & 1.37 & 0.107 & 0.0424 & 0.0397 \\ 
\hline
1.7 & -1.76 & -1.65 & -1.66  & -1.63 & -1.64 & 5.59 (339) & 0.995 & 0.805 & 0.0857 & 0.0362 & 0.0325 \\ 
\hline
1.8 & -1.77 & -1.69 & -1.70  & -1.67 & -1.68 & 4.55 (358) & 0.968 & 0.808 & 0.0771 & 0.0356 & 0.0325 \\ 
\hline
1.9 & -1.77 & -1.68 & -1.67  & -1.67 & -1.66  & 3.47 (356) & 0.774 & 0.552 & 0.0693 & 0.0328 & 0.0277 \\ 
\hline
2.0 & -1.77 & -1.71 & -1.71  & -1.70 & -1.70  & 3.02 (357)& 0.922 & 0.748 & 0.0642 & 0.0356 & 0.0320 \\ 
\hline
2.1 & -1.77 & -1.73 & -1.72  & -1.72 & -1.70  & 1.52 (222)& 0.277 & 0.224 & 0.0711 & 0.0308 & 0.0273 \\ 
\hline
2.2 & -1.77 & -1.76 & -1.75  & -1.74 & -1.73  & 1.36 (227) & 0.339 & 0.286 & 0.0669 & 0.0339 & 0.0307 \\ 
\hline
2.3 & -1.77 & -1.76 & -1.74  & -1.74 & -1.72  & 1.14 (228)& 0.306 & 0.250 & 0.0624 & 0.0345 & 0.0291 \\ 
\hline
2.4 & -1.77 & -1.76 & -1.74  & -1.74 & -1.73  & 1.02 (231)& 0.355 & 0.290 & 0.0577 & 0.0348 & 0.0307 \\ 
\hline
2.5 & -1.77 & -1.74 & -1.71  & -1.72 & -1.70  & 1.01 (239)& 0.356 & 0.299 & 0.0566 & 0.0345 & 0.0307 \\ 
\hline
2.6 & -1.77 & -1.74 & -1.71  & -1.73 & -1.70  & 0.955 (238) & 0.416 & 0.352 & 0.0560 & 0.0379 & 0.0338 \\ 
\hline
2.7 & -1.77 & -1.73 & -1.70  & -1.72 & -1.70 & 0.851 (242) & 0.430 & 0.351 & 0.0519 & 0.0378 & 0.0332 \\ 
\hline
2.8 & -1.75 & -1.59 & -1.52  & -1.60 & -1.56  & 1.36 (248) & 0.716 & 0.602 & 0.0646 & 0.0465 & 0.0427 \\ 
\hline
2.9 & -1.75 & -1.67 & -1.63  & -1.67 & -1.64  & 0.922 (248) & 0.512 & 0.420 & 0.0540 & 0.0413 & 0.0362 \\ 
\hline
3.0 & -1.75 & -1.69 & -1.65  & -1.69 & -1.66 & 1.14 (250) & 0.875 & 0.768 & 0.0574 & 0.0516 & 0.0470 \\ 
\hline
\end{tabular}
\caption{The parameters for each simulation for the $0.3M_{\odot}$, MAMS star: the parameters are the same as those in Table \ref{tab:1}.}
\label{tab:4}
\end{center}
\end{table*}

\end{document}